\documentclass[english,aps,prl,reprint,superscriptaddress,showpacs]{revtex4-1}

\usepackage[english]{babel}
\usepackage[utf8]{inputenc}				

\usepackage[cmex10]{amsmath}  			
\usepackage{amssymb}          			
\usepackage{amsfonts}         			

\usepackage{graphicx}					
\usepackage{subfigure}   
\usepackage[nolist]{acronym}

\usepackage[usenames,dvipsnames]{color}

\DeclareMathOperator*{\nRe}{Re}
\DeclareMathOperator*{\nIm}{Im}

\graphicspath{{figures/}}

\begin{document}

\title{Discontinuous Attractor Dimension at the Synchronization Transition of Time-Delayed Chaotic Systems}

\author{Steffen Zeeb}
\email{steffen.zeeb@physik.uni-wuerzburg.de}
\affiliation{Institute of Theoretical Physics, University of
  W\"urzburg, Am Hubland, 97074 W\"urzburg, Germany}

\author{Thomas Dahms}
\affiliation{Institute of Theoretical Physics, Technical University of
  Berlin, Hardenbergstra{\ss}e 36, 10623 Berlin, Germany}

\author{Valentin Flunkert}
\affiliation{Institute of Theoretical Physics, Technical University of
  Berlin, Hardenbergstra{\ss}e 36, 10623 Berlin, Germany}
\affiliation{Instituto de Fisica Interdisciplinar y Sistemas
  Complejos, IFISC (UIB-CSIC), Campus Universitat de les Illes
  Balears, E-07122 Palma de Mallorca, Spain}

\author{Eckehard Schöll}
\affiliation{Institute of Theoretical Physics, Technical University of
  Berlin, Hardenbergstra{\ss}e 36, 10623 Berlin, Germany}

\author{Ido Kanter}
\affiliation{Department of Physics, Bar-Ilan
  University, Ramat-Gan 52900, Israel}

\author{Wolfgang Kinzel}
\affiliation{Institute of Theoretical Physics, University of
  W\"urzburg, Am Hubland, 97074 W\"urzburg, Germany}

\date{\today}

\begin{abstract}
The attractor dimension at the transition to complete synchronization in a network of chaotic units with time-delayed couplings is investigated.
In particular, we determine the Kaplan-Yorke dimension from the spectrum of Lyapunov exponents for iterated maps and for two coupled semiconductor lasers. 
We argue that the Kaplan-Yorke dimension must be discontinuous at the transition and compare it to the correlation dimension.
For a system of Bernoulli maps we indeed find a jump in the correlation dimension. 
The magnitude of the discontinuity in the Kaplan-Yorke dimension is calculated for networks of Bernoulli units as a function of the network size.
Furthermore the scaling of the Kaplan-Yorke dimension as well as of the Kolmogorov entropy with system size and time delay is investigated.
\end{abstract}

\maketitle

\section{Introduction}

Networks of identical nonlinear units which are coupled by their dynamic variables can synchronize to a common chaotic trajectory \cite{Pecora-1990,Pikovsky1984}. This phenomenon is of fundamental interest in nonlinear dynamics, with applications in neural networks, coupled lasers, electronic networks and secure communication \cite{Boccaletti20021,Boccaletti2006175,RevModPhys.74.47,Arenas200893}.
For many applications, the coupling signals are transmitted with a time delay which is much larger than the internal time scales of the individual units. A network of identical units with identical delay times can synchronize to a common chaotic trajectory without any time shift between these units (zero lag synchronization) \cite{Klein2006}. This can only occur for so-called weak chaos \cite{PhysRevLett.107.234102}, i.e., if the largest \ac{LE} of the network decays with the inverse delay time.

A chaotic system with time delayed couplings (including self-feedback for a single unit) has special mathematical properties. The system becomes infinite dimensional and has a continuous spectrum of \acp{LE}. The dimension of the chaotic attractor increases proportional to the delay time \cite{Lepri-1994, Farmer-1982}. In this contribution we investigate the attractor dimension close to the transition to chaos synchronization.

The dimension of a chaotic attractor is not uniquely defined \cite{Farmer-1983}. We consider two attractor dimensions: The \ac{KY} dimension \cite{Kaplan-1979}, $D_{KY}$, which is determined from the spectrum of the \acp{LE}, and the correlation dimension, $D_C$ \cite{Grassberger1983189, PhysRevLett.50.346}. The \ac{KY} dimension is conjectured to be identical to the information dimension, $D_I$. This is known as the so-called \ac{KY} conjecture \cite{Frederickson-1983, Grassberger198434}. The information dimension in turn is an upper bound for the correlation dimension, $D_I \ge D_C$ \cite{Argyris}.

In this article we study these two attractor dimensions, $D_{KY}$ and $D_C$, for networks of identical nonlinear units coupled by their time-delayed variables. The delay time is much larger than the internal time scales, and chaos is generated by the coupling. Increasing the coupling strength, the system has a transition to complete synchronization \cite{PikovskyBook}. 

We show that the \ac{KY} dimension is discontinuous at this transition, it jumps from a high value for the unsynchronized to a low value for the synchronized chaotic attractor. This is a general results which holds for any networks with time-delayed couplings. We numerically calculate the \ac{KY} dimension for networks of iterated maps and for two coupled semiconductor lasers. For networks of Bernoulli maps we are also able to obtain analytical results. As a cross check the correlation dimension of the systems is computed as well and (at least) for networks of Bernoulli maps it displays a clear discontinuity at the synchronization transition in agreement with the jump in the \ac{KY} dimension. Finally we calculate the jump in the \ac{KY} dimension as a function of the size of the network.

\section{Discontinuous Kaplan-Yorke dimension}

Let us consider a network of $N$ identical nonlinear units. Each unit $j=1 , \ldots , N$ has a set of dynamic variables $x_j(t)$ which obey the following differential equations with time-delayed couplings

\begin{equation}
\label{net} 
\dot{x}_{j}(t) =F \! \left[x_j(t)\right] + \sigma \sum_{k=1}^{N} G_{jk} \, H \! \left[{x}_{k} \! \left(t-\tau \right) \right] \: .
\end{equation}

The function $F$ describes the local dynamics of each unit, the function $H$ couples the time-delayed variables of the connected units and the adjacency matrix $G_{jk}$ defines the graph of the network. We restrict this matrix to have only nonnegative entries and a constant row sum $\sum_k G_{jk}=1$. 
Thus the eigenvalue of $G$ with the largest modulus is always $\gamma_1=1$.
The parameter $\sigma$ determines the strength of the coupling and $\tau$ is the delay time of the coupling.

By construction, the \ac{SM} $x_j(t)=s(t)$ is a solution of this network given by the equation

\begin{equation}
\label{sm} \dot{{s}}(t) ={F}\!\left[{s}(t)\right] + \sigma {H}\!\left[{s}\!\left(t-\tau \right)\right] \: .
\end{equation}
We consider only networks where this equation has chaotic solutions for sufficiently large values of $\sigma$, i.e., the dynamics on the \ac{SM} is chaotic and has at least one positive \ac{LE}. The stability of the \ac{SM} can be determined using the master stability function \cite{Pecora-1998}. An infinitesimal perturbation of the synchronized trajectory can be decomposed into eigenvectors of the coupling matrix $G$ with corresponding eigenvalues $\gamma_k$, $k=1,\ldots,N$. The amplitudes $\xi_k(t)$ of the perturbations along these eigenvectors follow the equation
\begin{equation}
\label{msf} \dot{\xi}_k(t)=DF[s(t)]\, \xi_k(t) + \sigma \gamma_k DH[s(t-\tau)]\, \xi_k(t-\tau)  \: .
\end{equation}
Hence the perturbations are governed by a linear differential equation with time-delayed feedback and time dependent coefficients. For each eigenvalue $\gamma_k$ of $G$ this equation yields a whole spectrum of \acp{LE}. 
A perturbation corresponding to the largest eigenvalue, $\gamma_1=1$, describes a perturbation within the \ac{SM}. Hence $\gamma_1$ is called transversal eigenvalue. In the chaotic regime, the linearized dynamics is unstable for $\gamma_1=1$, i.e., it has positive \acp{LE}. Any arbitrary perturbation has in general components in the other eigenmodes with eigenvalues $\gamma_k, k=2,...,N $. These are the so-called longitudinal eigenvalues. Thus the \ac{SM} is stable if eq.~\eqref{msf} yields only negative \acp{LE} for all transversal eigenvalues such that perturbations transversal to the \ac{SM} are decaying exponentially fast.


We consider only networks where the delay time $\tau$ is much larger than any other time scale of the system \cite{PhysRevLett.105.254101}. In this case the condition for stability of the \ac{SM} is given by the following equation \cite{PhysRevLett.107.234102}

\begin{equation}
\label{gap} |\gamma_2| < \exp(- \lambda_m \tau)  \: .
\end{equation}

Hence the transversal eigenvalue of $G$ with the largest absolute
value, $|\gamma_2|$, and the maximum (longitudinal) \ac{LE},
$\lambda_m$, of a single unit with feedback, given by eq.~\eqref{sm},
determine the stability of complete chaos synchronization.
Note that $\lambda_m$ depends on the coupling strength $\sigma$,
therefore eq.~\eqref{gap} determines the critical coupling strength
$\sigma_c$ where chaos synchronization appears.

In this work, we change a control parameter of the system such that the system exhibits a transition from from a synchronized to an unsynchronized state. At this transition we investigate the change in the attractor dimension.

A quantitative measure for the structure of the attractor is the \ac{KY} dimension. It is defined by the spectrum of \acp{LE} obtained from eq.~\eqref{msf}. Considering a discrete spectrum $\lambda_1 \ge \lambda _2 \ge \ldots $, the \ac{KY} dimension is defined as the largest number $M$ for which the sum of LEs is still positive plus an interpolation term which yields the fractal part of the dimension,

\begin{equation}
\label{EqnKaplanYorke} D_{KY}=M + \frac{\sum_{k=1}^M \lambda_k}{|\lambda_{M+1}|}  \: .
\end{equation}
  
Now we want to show that the attractor dimension at the
transition to chaos synchronization is discontinuous. The \acp{LE} are given by
eq.~\eqref{msf} where one obtains a whole spectrum of \ac{LE} for each
eigenvalue $\gamma_k$.

\begin{figure}
	\centering
    \includegraphics[width=0.95\linewidth]{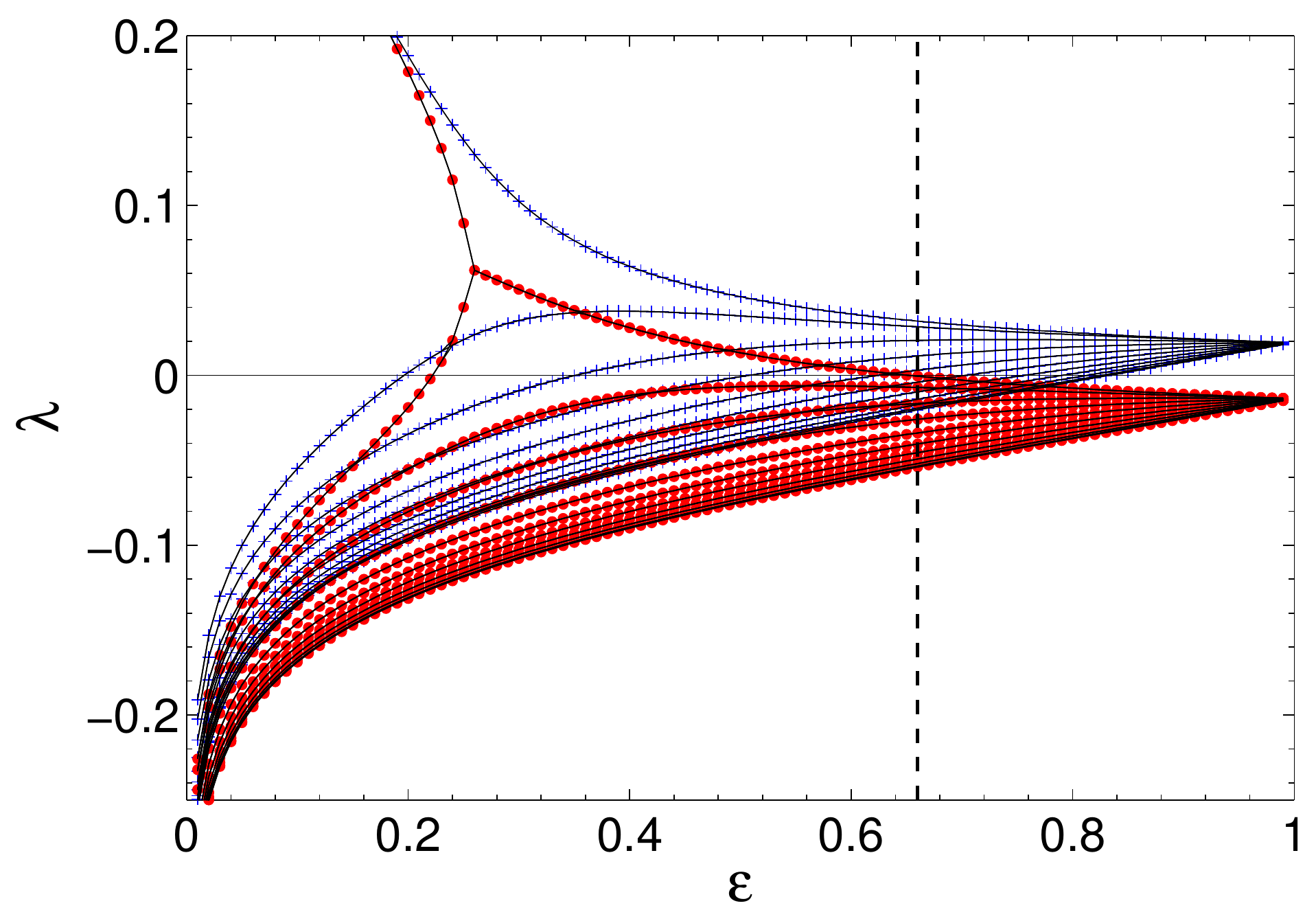}
	\caption{(Color online) Lyapunov spectra vs.\ coupling strength $\epsilon$ for a
      system of three all-to-all coupled Bernoulli maps with the
      parameters $a=1.5$ and $\tau=20$. The blue crosses show the
      longitudinal spectrum associated with $\gamma_1=1$ and the red
      dots show the transversal spectrum associated with
      $\gamma_2=\gamma_3=-1/2$. Both spectra were obtained by solving the
      polynomial equations derived from the master stability function.
      The black solid lines show the Lyapunov spectrum computed by Gram-Schmidt from simulations of the full system. The vertical dashed line
      indicates $\epsilon_c$ where to its right the system is
      synchronized.}
	\label{FigLyapSpecEps}
\end{figure}

First, consider stable synchronization. For the longitudinal
eigenvalue $\gamma_1=1$ the perturbations have in general positive as
well as negative \acp{LE}. For all other (i.e., transversal)
eigenvalues the corresponding \acp{LE} are negative since the \ac{SM}
is stable. But close to the synchronization transition the maximum
\ac{LE} of (at least) one of these transversal Lyapunov spectra
approaches the value zero, as shown in Fig.~\ref{FigLyapSpecEps}.
Consequently, close to the transition these spectra contribute to the
definition of the \ac{KY} dimension, defined by
eq.~\eqref{EqnKaplanYorke}.

However, this cannot be true. In case of stable synchronization the trajectory of the network, determined by eq. \eqref{sm}, is completely restricted to the \ac{SM}. Two neighboring trajectories inside the \ac{SM} deviate from each other according to eq.~\eqref{msf} with $\gamma_1=1$. Any transversal eigenvalues cannot contribute to the dynamics inside the \ac{SM}. Thus, only the longitudinal \ac{LE} spectrum can contribute to the KY dimension, i.e., the sum in eq.~\eqref{EqnKaplanYorke} must only run over the \acp{LE} of the $\gamma_1$ spectrum.

Now, consider the dynamics outside but still close to the synchronization transition. In this case, the perturbations cannot be decomposed into the eigenmodes of the coupling matrix $G$ since the coefficients of the linearized equations of eq.~\eqref{net} depend on each single node. But close to a supercritical transition we expect that the \acp{LE} are continuous as a function of $\sigma$ and the structure of the \acp{LE} is still similar to the one inside the \ac{SM}, as it is the case in Fig.~\ref{FigLyapSpecEps}.

In the desynchronized regime all eigenvalues contribute to the \ac{KY} dimension. Consequently, the \ac{KY} dimension must be discontinuous at the transition to chaos synchronization. 

Chaos synchronization can only occur for weak chaos, where the largest \ac{LE} scales with $1/\tau$. In this case the spectrum of \acp{LE} is dense. Hence, a large number of additional \acp{LE} contributes to the \ac{KY} dimension at the transition, and we expect the jump of the \ac{KY} dimension to be of the order of the size, $N$, of the network.

From the Lyapunov spectrum we can also calculate the Kolmogorov entropy which quantifies the predictability of the system \cite{SchusterBook}. It is defined as the sum over all positive \acp{LE}
\begin{equation}
	K = \sum_i{\lambda_i} \, , \quad \mathrm{for} \, \lambda_i > 0 \: .
	\label{EqnKolmogorov}
\end{equation}
Since only positive \acp{LE} contribute, the entropy is always defined with the complete spectrum of \acp{LE}, hence it does not show a jump at the transition to chaos synchronization. Nevertheless, at the transition at least one band of \acp{LE} adds to the entropy and we expect a discontinuous derivative of $K(\sigma)$

The jump in the dimension of the chaotic attractor and the kink in the entropy are general results which should hold for any chaotic network at the transition to chaos synchronization. In the following section we calculate the attractor dimensions $D_{KY}$ and $D_C$ and the entropy $K$ for networks of iterated maps.

\section{Iterated Maps}

The previous general statement holds not only for differential equations but also for networks of iterated maps with time-delayed coupling. Since such models are easier to analyze than continuous systems we investigate networks of iterated maps in this section. Each unit $j$ has a one-dimensional variable in the unit interval $x^j_t \in [0,1]$ which is updated in discrete time steps $t$ according to the following equation

\begin{equation}
\label{d-net}
x_{t+1}^{j}=(1-\epsilon)f(x_{t}^{j})+\epsilon \sum_{k}G_{jk}f(x_{t-\tau}^{k}) \: .
\end{equation}

The parameter $\epsilon$ is the coupling strength, but since the dynamic variable should stay in the unit interval, we subtract the undelayed term with strength $\epsilon$. For the function $f(x)$ we use the Bernoulli shift and the asymmetric tent map,

	\emph{Bernoulli map} 
	\begin{align*}
		x_{t+1} = (a \, x_t) \mod 1
	\end{align*} 
	
	\emph{Tent map}
	\begin{align*}
		x_{t+1} =   \left\{ 
		\begin{array}{ll} 
			\frac{1}{a} \, x_t & \mathrm{for} \: 0 \leq x_t < a \\
			\frac{1}{1-a} \, (1-x_t) & \mathrm{for} \: a \leq x_t \leq 1
		\end{array} \right. 
	\end{align*}
				
The Bernoulli map is chaotic for a parameter $a>1$. For the chaotic tent map the parameter $a$ is chosen such that the value $x_{t+1}$ stays in the range $[0,1]$, i.e., $0<a<1$. For maps, we have a discrete delay time $\tau$. As before, the coupling matrix $G$ has constant row sum $\sum_k G_{jk}=1$. 

The synchronized state is a solution of these equations and reads as

\begin{equation}
\label{d-syn}
s_{t+1}=(1-\epsilon) f(s_{t}) + \epsilon f(s_{t-\tau}) \: .
\end{equation}

As before, the perturbations of the \ac{SM} can be associated with the eigenvalues $\gamma_k$ of the coupling matrix $G$. The amplitude $\xi_t^k$ of each mode obey the linear master stability function determined by

\begin{equation}
\label{d-msf}
\xi_{t+1}^k = (1-\epsilon) f'(s_{t}) \, \xi_{t}^k + \epsilon \, \gamma_k f'(s_{t-\tau}) \, \xi_{t-\tau}^k \: .
\end{equation}

Note that for the Bernoulli network the derivative $f'=a$ is constant, therefore one only has to analyze linear equations with constant coefficients. For the tent map, however, the derivative can take on two different values and, hence, the coefficients change with time.

Since $\tau$ is discrete, for each mode with eigenvalue $\gamma_k$ one obtains $\tau+1$ many \acp{LE}. Fig.~\ref{FigLyapSpecEps} shows an example for a triangle of all-to-all coupled Bernoulli units with $\gamma_1=1$ and $\gamma_2=\gamma_3=-1/2$. For all values of $\epsilon$ the system is chaotic, since the largest \ac{LE} of the $\gamma_1$ band is always positive. The transition to chaos synchronization occurs at $\epsilon_c$ where the maximum \ac{LE} of the $\gamma_{2,3}$ band crosses the value zero.

The Lyapunov spectrum is in general obtained from a Gram-Schmidt orthonormalization procedure according to Farmer \cite{Farmer-1982}. The system's equation~\eqref{net} and \eqref{d-net}, respectively, are linearized around the chaotic trajectory and simulated for a set of orthogonal perturbation vectors which have to be re-orthogonalized after an appropriate amount of time. The Lyapunov spectrum is computed from the change in magnitude of the perturbation vectors. 

For Bernoulli networks we can also derive a polynomial equation of degree $\tau+1$ for the different eigenvalues $\gamma_k$ of the adjacency matrix $G$ from which the Lyapunov spectrum can easily be calculated. The polynomial equation reads as follows
\begin{align}
	z^{\tau+1} = (1- \epsilon) \, a \, z^\tau + \epsilon \, a \, \gamma_k  \: ,
	\label{PolynomialEqn}
\end{align}
where the \acp{LE} are given by $\lambda = \ln{|z|}$ \cite{PhysRevE.83.046222}. This equation still holds in the desynchronized region since it does not depend on the trajectory of the system. Thus the \ac{LE} spectrum can be calculated from eq.~\eqref{d-msf} and \eqref{PolynomialEqn}, respectively, in the complete parameter space. Note that both methods -- orthonormalization procedure and polynomial equation -- which compute the spectrum in completely different ways yield the same results, see Fig.~\ref{FigLyapSpecEps}. 

From Figs.~\ref{FigLyapSpecEps} and \ref{FigLyapSpecEpsTent} it can be seen that for $\epsilon$-values close to $1$ the \acp{LE} cluster into bands.
This can be understood as follows.
For $\epsilon=1$, the dynamical equations~\eqref{d-net} are given by
\begin{equation} 
  x_{t+1}^j =  \sum_{k=1}^N G_{jk} f(x_{t-\tau}^k)  \;.  
\end{equation}
Since the state at time $t+1$ is only influenced by the state at time $t-\tau$, the system is effectively given by $\tau+1$ uncoupled identical systems of the form
\begin{equation} 
  \tilde{x}^j_{\theta+1}=\sum_{k} G_{jk} f(\tilde{x}_\theta^k) \;.
  \label{eq:effective}
\end{equation} 
Each of these $\tau+1$ systems is $N$-dimensional and gives rise to $N$ Lyapunov exponents.
Since we have $\tau+1$ identical systems, each of these $N$ exponents is $\tau+1$ times degenerate.
This holds as long as each effective system evolves on the same chaotic attractor, and thus does not rely on synchronization.

For $\epsilon<1$ the first term in eq.~\eqref{d-net} leads to a coupling between these effective systems and thus removes the degeneracy.


It is insightful to discuss this lifting of degeneracy for the case of the Bernoulli maps.
For $\epsilon=1$ the solutions of the variational eq.~\eqref{PolynomialEqn} are given by the complex $(\tau+1)$-th roots of $\gamma_k a$
\begin{equation} 
  z^{(0)} = (\gamma_k a)^{\frac{1}{\tau+1}}\, e^{i\frac{2\pi}{\tau+1} l}  \qquad (l=0,\dots,\tau) \;,  
\end{equation}
and the corresponding $(\tau+1)$ Lyapunov exponents are all equal and are given by
\begin{equation} 
  \lambda^{(0)} = \ln|z^{(0)}| = \ln|\gamma_k a|^{\frac{1}{\tau+1}}\;.
\end{equation}
Here the superscript $(0)$ indicates the zero-th order in an expansion in $1-\epsilon$.

\begin{figure}
  \centering
  \includegraphics{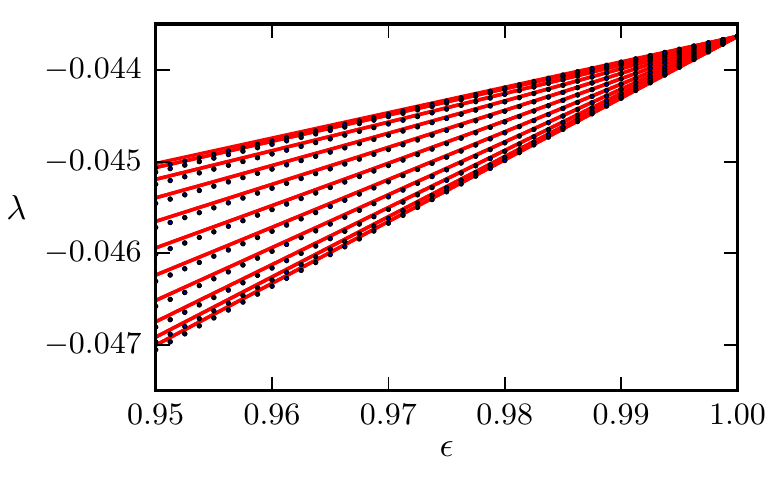}
  \caption{\label{fig:perturbation}(Color online) Perturbation expansion of
    eq.~\eqref{PolynomialEqn} up to first order in $\mu=(1-\epsilon)$
    (red lines) and exact location of eigenvalues (blue dots).
    Parameters: $a=0.4$, $\gamma_k=1$, $\tau=20$.}
\end{figure}

\begin{figure}
	\centering
    \includegraphics[width=0.95\linewidth]{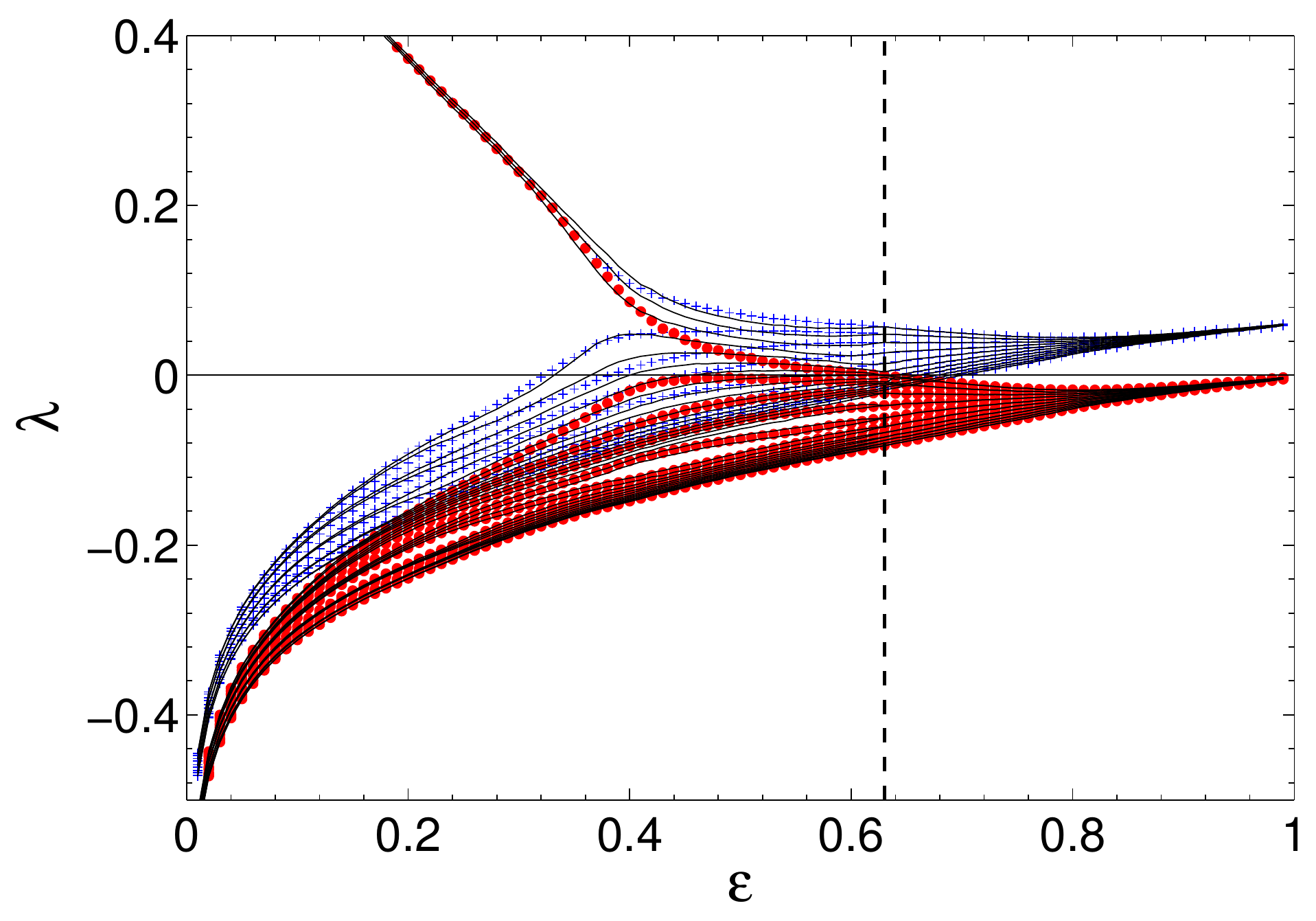}
	\caption{(Color online) Lyapunov spectra vs.\ coupling strength $\epsilon$ for a
      system of three all-to-all coupled Tent maps with the parameters
      $a=0.4$ and $\tau=10$. The blue crosses show the longitudinal
      spectrum associated with $\gamma_1=1$ and the red dots show the
      transversal spectrum associated with $\gamma_2=\gamma_3=-1/2$
      computed by Gram-Schmidt from simulations of the master stability function.
      The black solid lines show the Lyapunov spectrum computed by Gram-Schmidt from simulations of the full system.
      The vertical dashed line indicates $\epsilon_c$ where to its
      right the system is synchronized. Note that for a better
      visibility a system with $\tau=10$ is plotted. 
 }
	\label{FigLyapSpecEpsTent}
\end{figure}

For $\epsilon \ne 1$, we can make a perturbation expansion of eq.~\eqref{PolynomialEqn} in the small parameter $\mu=1-\epsilon$.
Inserting the ansatz $z = z^{(0)} + \mu \beta^{(1)}$, where $\beta^{(1)}$ is a coefficient that needs to be determined, into eq.~\eqref{PolynomialEqn} yields the solutions up to first order in in $\mu$
\begin{equation} 
  z^{(1)} = z^{(0)} + \mu \beta^{(1)} = z^{(0)} + \mu \frac{a}{\tau+1}\left(1-\frac{\gamma_k}{\left(z^{(0)}\right)^\tau}\right) \;.
\end{equation}
The values of the corresponding Lyapunov exponents ($\lambda^{(1)}=\ln|z^{(1)}|$) are depicted in Fig.~\ref{fig:perturbation} as a function of $\epsilon$.
One clearly sees the lifting of degeneracy due to $\epsilon < 1$.

 Although for general maps $f$ one cannot write down an
equation such as \eqref{PolynomialEqn} in the unsynchronized regime,
for $\epsilon=1$ the degeneracy follows rigorously from the discussion
above and for $\epsilon<1$ the interaction will generically lead to a
lifting of degeneracy similar to that shown in
Fig.~\ref{fig:perturbation}.

In the synchronized regime the linear equations \eqref{msf} and
\eqref{d-msf} can be used to compute the \ac{LE} spectrum for any
iterated map $f(x)$. Hence the \acp{LE} can be clustered into bands
according to the eigenvalues of the coupling matrix $G$. However, for
the unsynchronized system these linear equations only hold for
Bernoulli networks where the coefficients are constant and in
particular independent of the systems trajectory. In general, in the
desynchronized regime we need to evaluate the linearized equations of
the full system, eq. \eqref{net} and \eqref{d-net}, to obtain the
\ac{LE} spectrum, and cannot restrict to the master stability
function, eq. \eqref{msf} and \eqref{d-msf}. But close to the
transition we expect the spectra obtained from the master stability
function to approximate the true spectra very well. Surprisingly, for
the tent map, the results coincide very well not only close to the
synchronizations transition but for all values of $\epsilon$ when for
the desynchronized system in the master stability function, eq.
\eqref{d-msf}, the dynamics of a single unit is inserted, e.g.,
$s_{t}$ and $s_{t-\tau}$ is replaced by $x_{t}^{1}$ and
$x_{t-\tau}^{1}$, respectively. A comparison of the different spectra
is shown in Fig.~\ref{FigLyapSpecEpsTent}. In contrast to to
Fig.~\ref{FigLyapSpecEps}, the blue and red lines in
Fig.~\ref{FigLyapSpecEpsTent} are obtained using the Gram-Schmidt
procedure from simulating the master stability function which is
strictly only valid for the synchronized regime. The black line is
obtained using the Gram-Schmidt procedure on the full system's
equations and therefor yields the correct results not only for the
synchronized but also for the unsynchronized regime. Within the
synchronized regime the results match up to numerical accuracy,
whereas outside of synchronization the results of the two methods
deviate since the master equation is no longer valid.

\begin{figure}
	\centering
		\includegraphics[width=6.5cm]{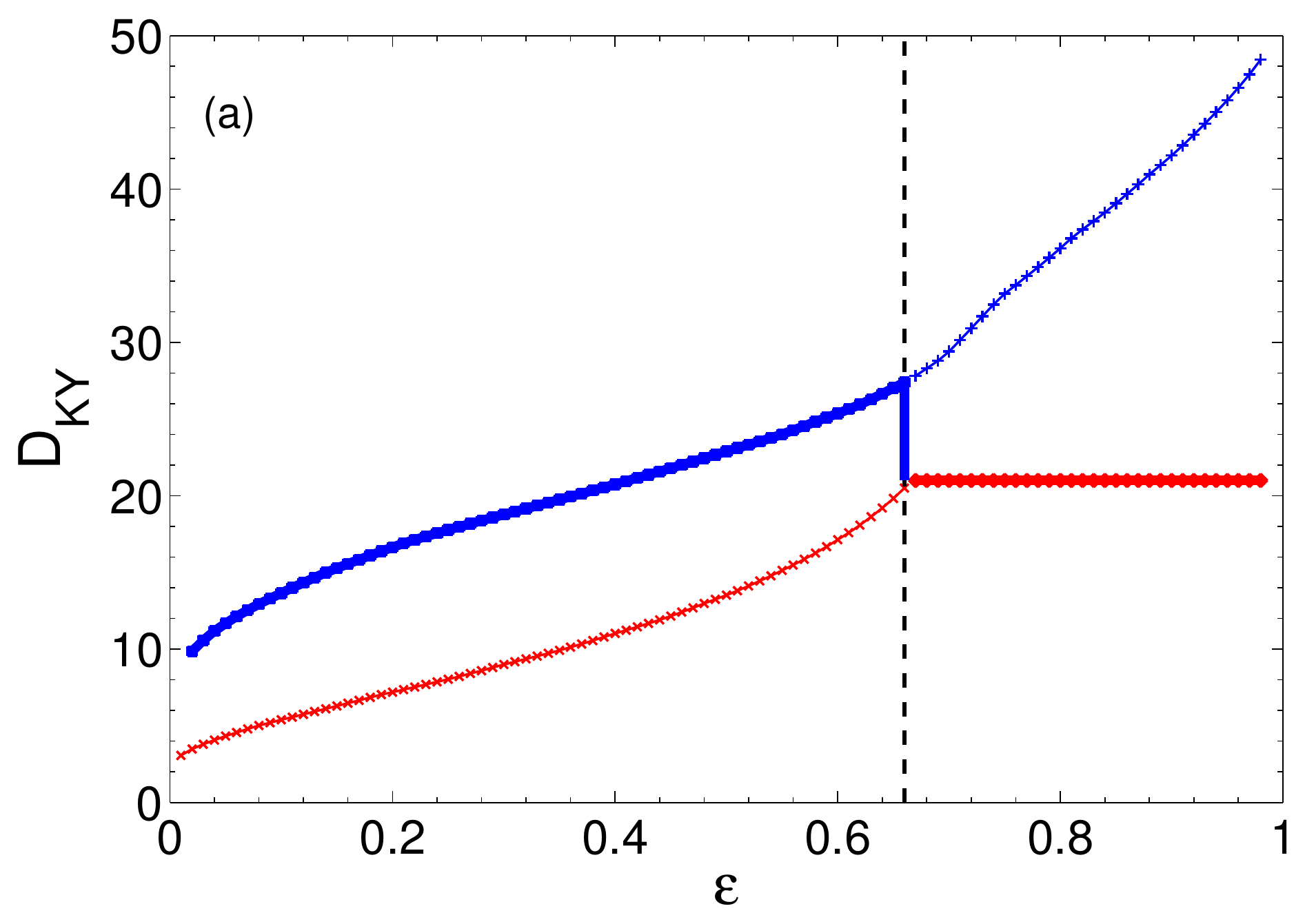}
		\includegraphics[width=6.5cm]{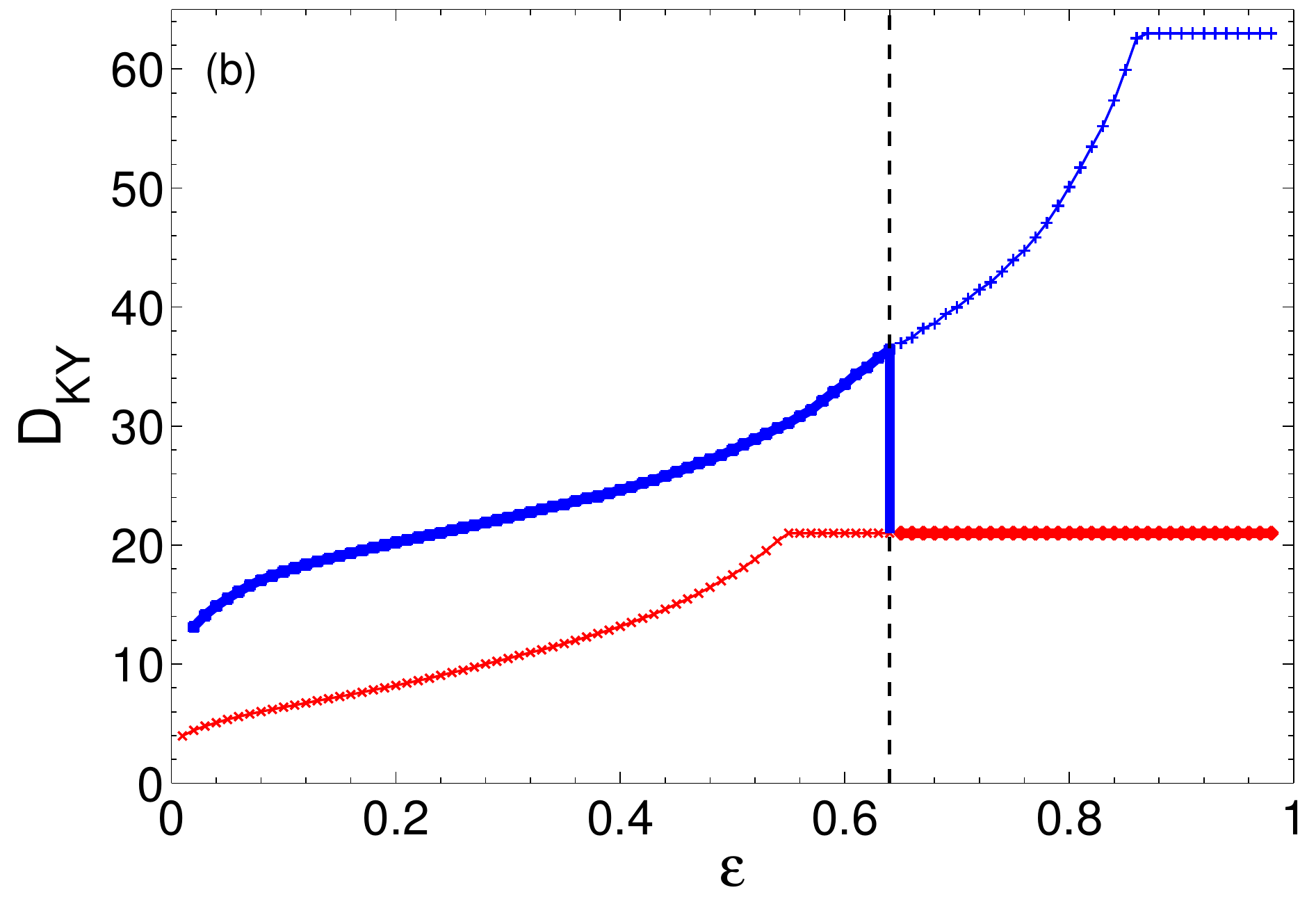}
	\caption{(Color online) \ac{KY} dimension $D_{KY}$ with respect to the coupling strength $\epsilon$ for a system of three all-to-all coupled (a) Bernoulli and (b) tent maps, respectively. The upper blue line shows $D_{KY}$ of the full system, the lower red line shows $D_{KY}$ of the \ac{SM}. The vertical dashed line indicates $\epsilon_c$ where to its right the system is synchronized. The parameters are $\tau=20$, $a=1.5$ for the Bernoulli map and $a=0.4$ for the tent map, respectively.}
	\label{Fig3UnitsKYDimEps}
\end{figure}

The \ac{KY} dimension for the triangle of Bernoulli and tent maps, respectively, is shown in Fig.~\ref{Fig3UnitsKYDimEps}. The upper curve shows the \ac{KY} dimension when the complete set of \acp{LE} is used in eq.~\eqref{EqnKaplanYorke}, while the lower curve uses only the \acp{LE} of the \ac{SM} which are obtained from simulating a single unit. Note that an upper bound of the dimension is $3(\tau+1)$, i.e., the full system's dimension, for the desynchronized triangle and $\tau+1$ for the manifold. 

As discussed in the previous section, in the synchronized region the lower curve is valid whereas in the desynchronized region the upper curve is valid. Thus the \ac{KY} dimension jumps to a lower value when the parameter $\epsilon$ is increased above $\epsilon_c$.

In the following we consider a pair of maps with self-feedback. The dynamic equations of the system read
\begin{equation}
	x_{t+1}^i = (1-\epsilon) f(x_t^i) + \epsilon \kappa f(x_{t-\tau}^i) + \epsilon (1-\kappa) f(x_{t-\tau}^j) \: ,
	\label{Eqn-d-pair}
\end{equation}
with $i,j \in \left\{1,2\right\}$. The parameter $\epsilon$ is, as before, the coupling strength of the delayed terms to the internal dynamics and the parameter $\kappa$ determines the ratio between the external coupling and the self-feedback. The synchronized trajectory of the system also follows eq.~\eqref{d-syn} which does not
contain the strength $\kappa$ of the self-feedback. Hence the synchronized trajectory, i.e., the \ac{SM} is independent of $\kappa$ and only changes with the coupling strength $\epsilon$. Fig.~\ref{Fig2UnitsKYDimKap} shows the \ac{KY} dimension as a function of $\kappa$. At the transition the dimension jumps from the upper (blue) curve to the lower (red) constant value. 

According to the discussion of the previous section the \ac{KY} dimension has to jump at the synchronization transition. This qualitative prediction of the \ac{KY} conjecture should be valid for any measure of the dimension of the chaotic attractor. Thus, we also computed the correlation dimension of the system to compare it to the \ac{KY} dimension and to check whether the dimension indeed jumps as we argue. 
For this purpose we analyzed the system's trajectories, i.e., the time series of the system using the TISEAN package of Kantz and Schreiber \cite{TISEAN}. In particular, the correlation function $C(\xi)$ according to Grassberger \& Procaccia was computed which scales as a power law $C(\xi) \propto \xi^{D_C}$ with the exponent being the correlation dimension \cite{Grassberger1983189, PhysRevLett.50.346}. 
A straight line was fitted to different correlation functions of different embedding dimensions in a double-logarithmic plot and at the same time the results were cross-checked in plots of the local slopes of the correlation function $d(\xi)=\partial C(\xi) / \partial \xi$ in which the power law behavior corresponds to a plateau. For more details on how to actually compute the correlation dimension the reader is referred to \cite{TISEAN, KantzSchreiber}. Note that this method allows a reliable calculation of the correlation dimension for small values of the delay $\tau$, only.  

\begin{figure}
	\centering
        \includegraphics[width=7cm]{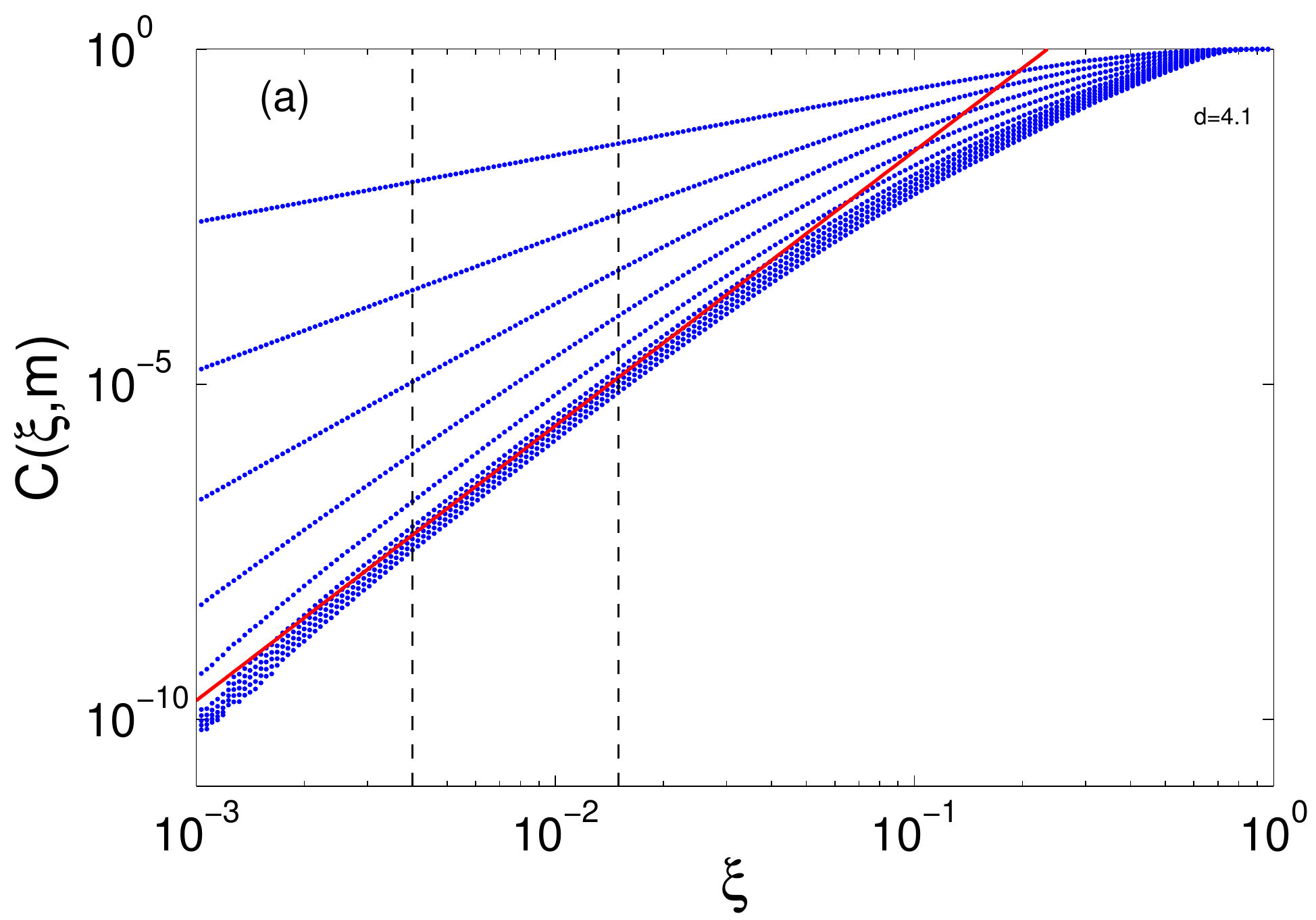}    
        \includegraphics[width=7cm]{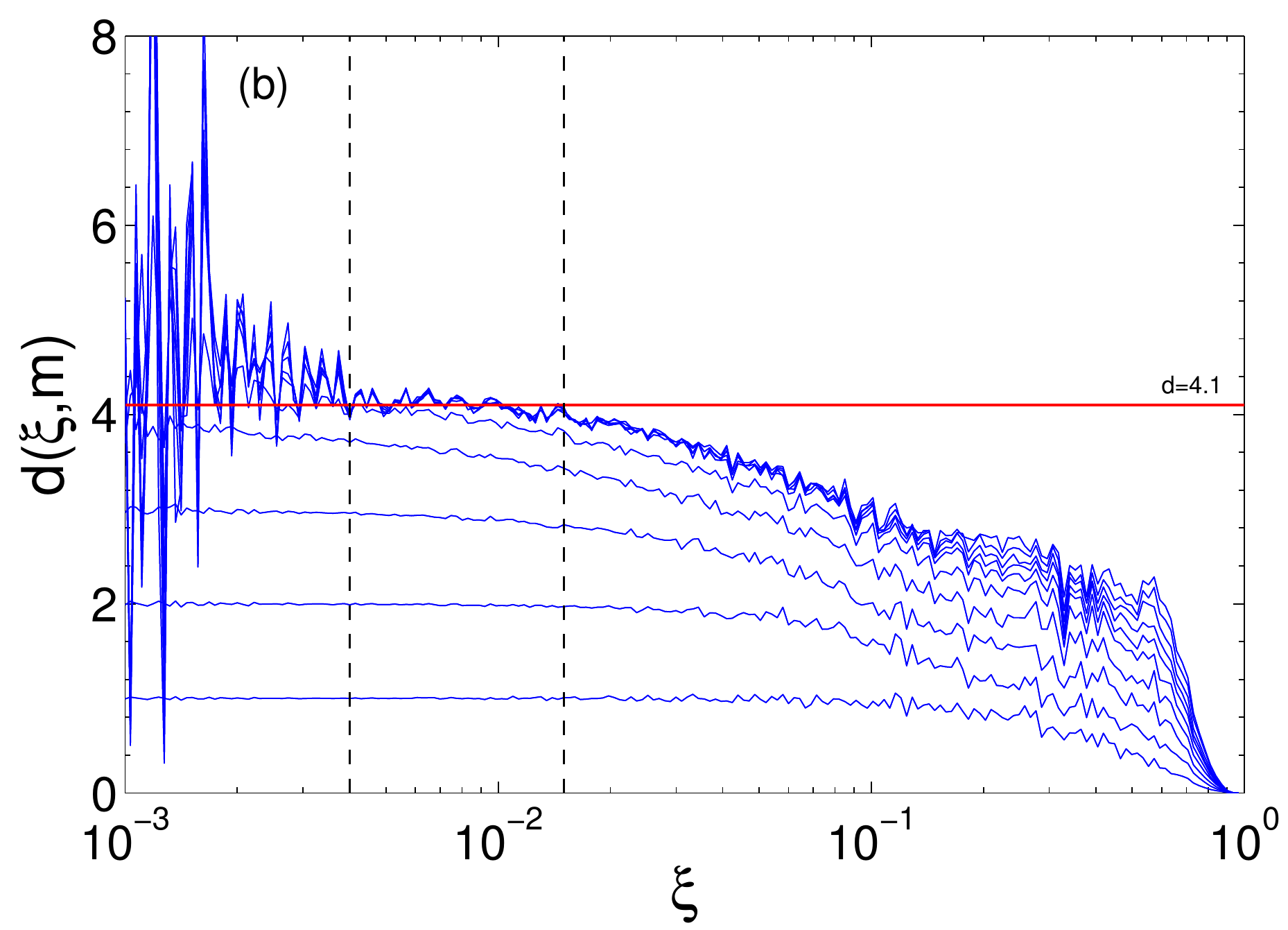}
	\caption{(a) Correlation function $C(\xi)$ and (b) local slope $d(\xi)$ for different embedding dimensions $m$ (different curves) computed from a timeseries of a system of two mutually coupled tent maps. The time delay is $\tau=5$, the length of the time series is $l=10^6$ and the other parameters are $a=0.4$, $\epsilon=0.45$ and $\kappa=0.35$. Vertical dashed lines indicate the range which was used for the fit.}
	\label{Fig2CorrFunc}
\end{figure}

\begin{figure}
	\centering
        \includegraphics[width=7cm]{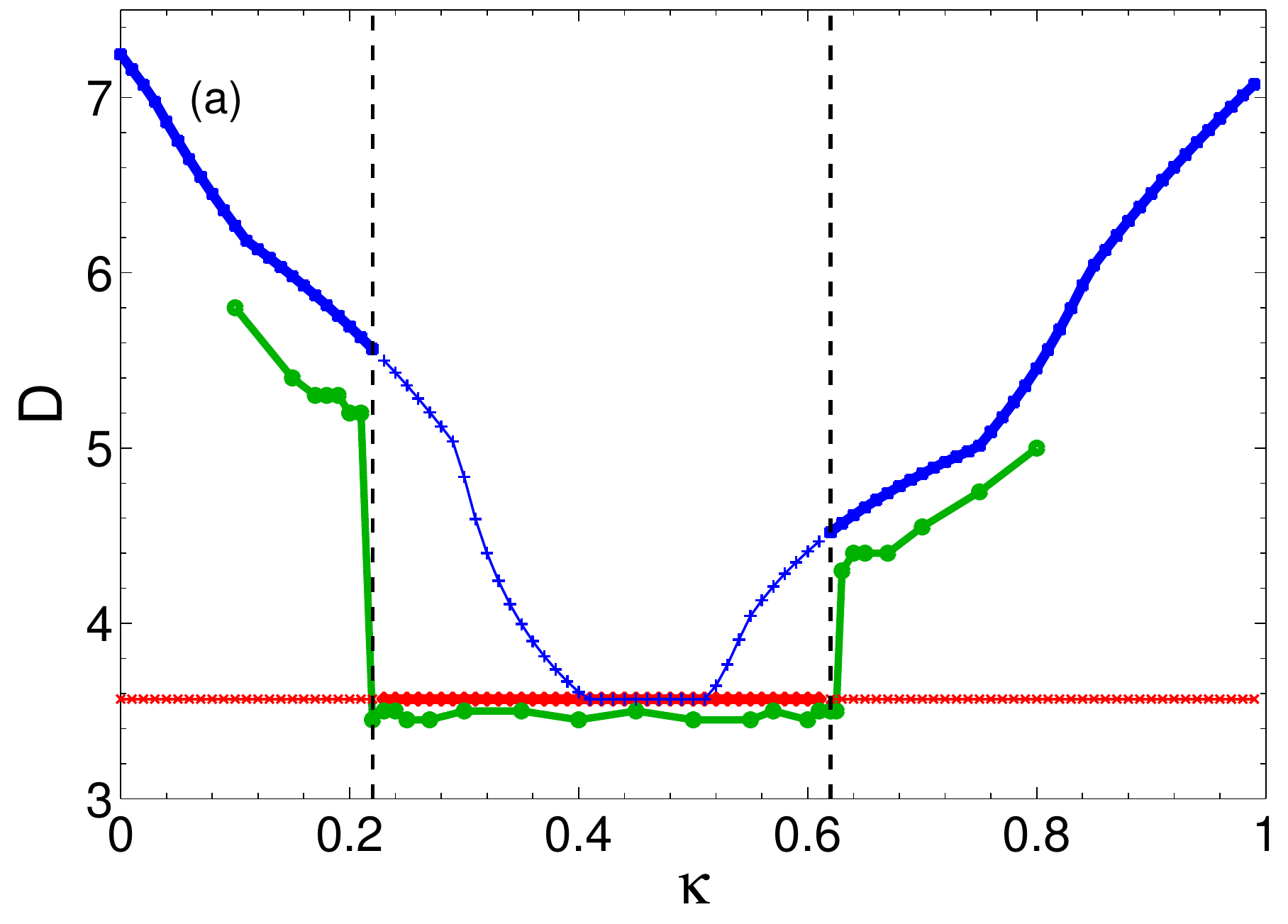}    
        \includegraphics[width=7cm]{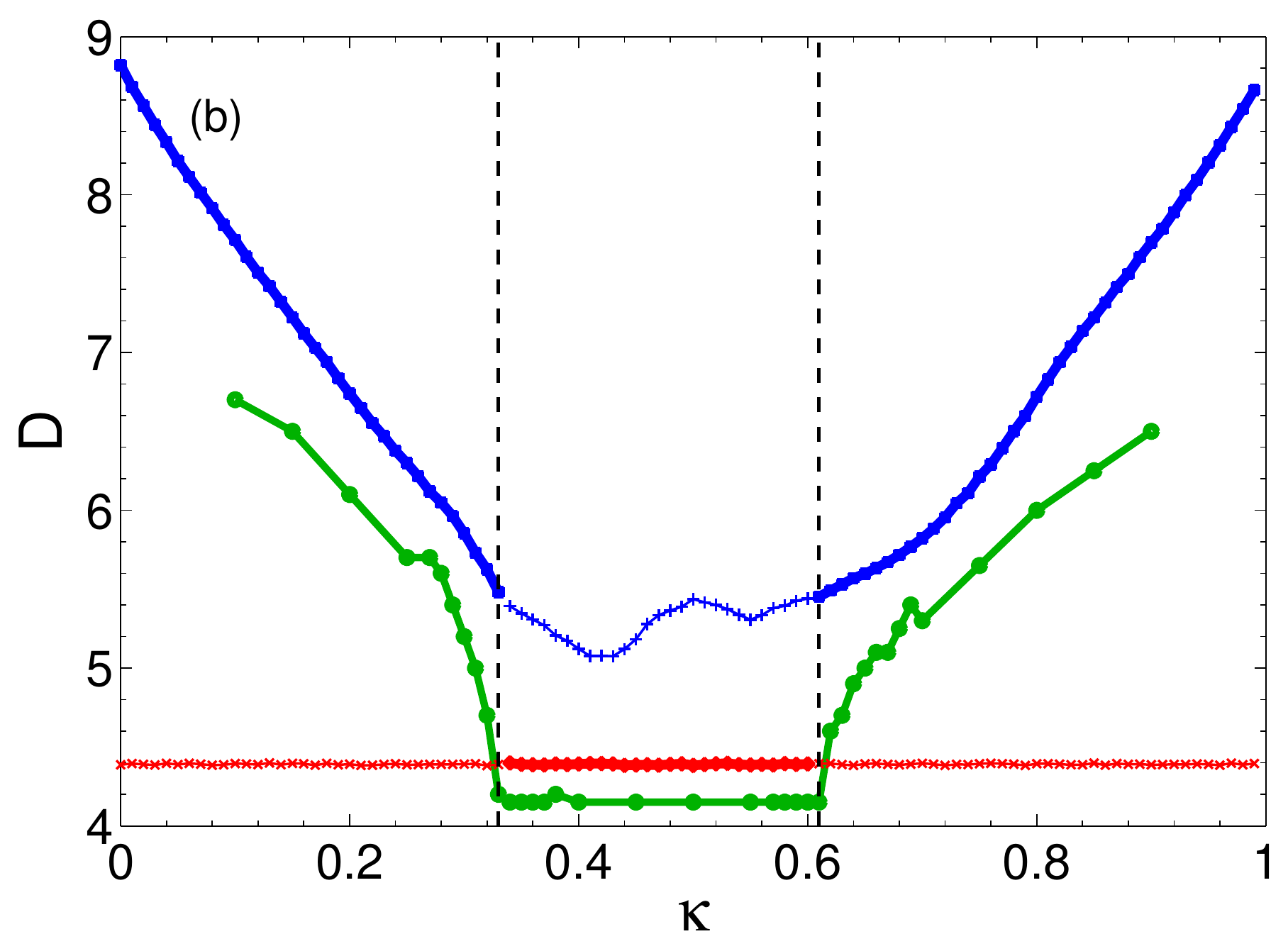}
	\caption{(Color online) Attractor dimension vs.\ coupling strength $\kappa$ for a system of two mutually coupled (a) Bernoulli and (b) tent maps, respectively. The time delay is $\tau=5$. The upper blue line shows $D_{KY}$ of the full system, the lower red line shows $D_{KY}$ of the \ac{SM} corresponding to $\gamma_1$ and the green curve in between shows $D_{C}$. The vertical dashed lines indicate $\kappa_c$. The parameters are $\tau=5$, $\epsilon=0.45$, $a=1.5$ for the Bernoulli map and $a=0.4$ for the tent map, respectively.}
	\label{Fig2UnitsKYDimKap}
\end{figure}

A typical plot of the correlation functions as well as the local slopes is shown in Fig.~\ref{Fig2CorrFunc}. This figure shows that the extrapolation of the slope to low values of the radius $\xi$ is difficult. Hence our values for the correlation dimension give a lower bound to the actual correlation dimension since due to the limited computational power it was not possible to analyze very small values of the distance $\xi$ with an appropriate accuracy. 


Fig.~\ref{Fig2UnitsKYDimKap} shows the results for the \ac{KY} and correlation dimension as a function of $\kappa$. The \ac{KY} dimension is larger than the correlation dimension in agreement with known theoretical inequalities. For the Bernoulli map the correlation dimension displays a clear jump at the transition to synchronization, in agreement with our theoretical prediction. For the tent map, however, the discontinuity is not clearly visible from our results. As stated above, for small distances $\xi$ the correlation function $C(\xi)$ shows large fluctuations due to the limited statistics. From our results of Fig.~\ref{Fig2CorrFunc} we cannot rule out that the local slopes might not be saturated yet. Thus the obtained results only serve as a lower bound which, according to our results, seems to increase with longer trajectories and better statistics. Consequently, the results of Fig.~\ref{Fig2UnitsKYDimKap} do not rule out a discontinuous behavior of the attractor dimension. In any case, the synchronization transition is clearly visible in the discontinuous slope of the correlation dimension.

\begin{figure}
	\centering
        \includegraphics[width=6.5cm]{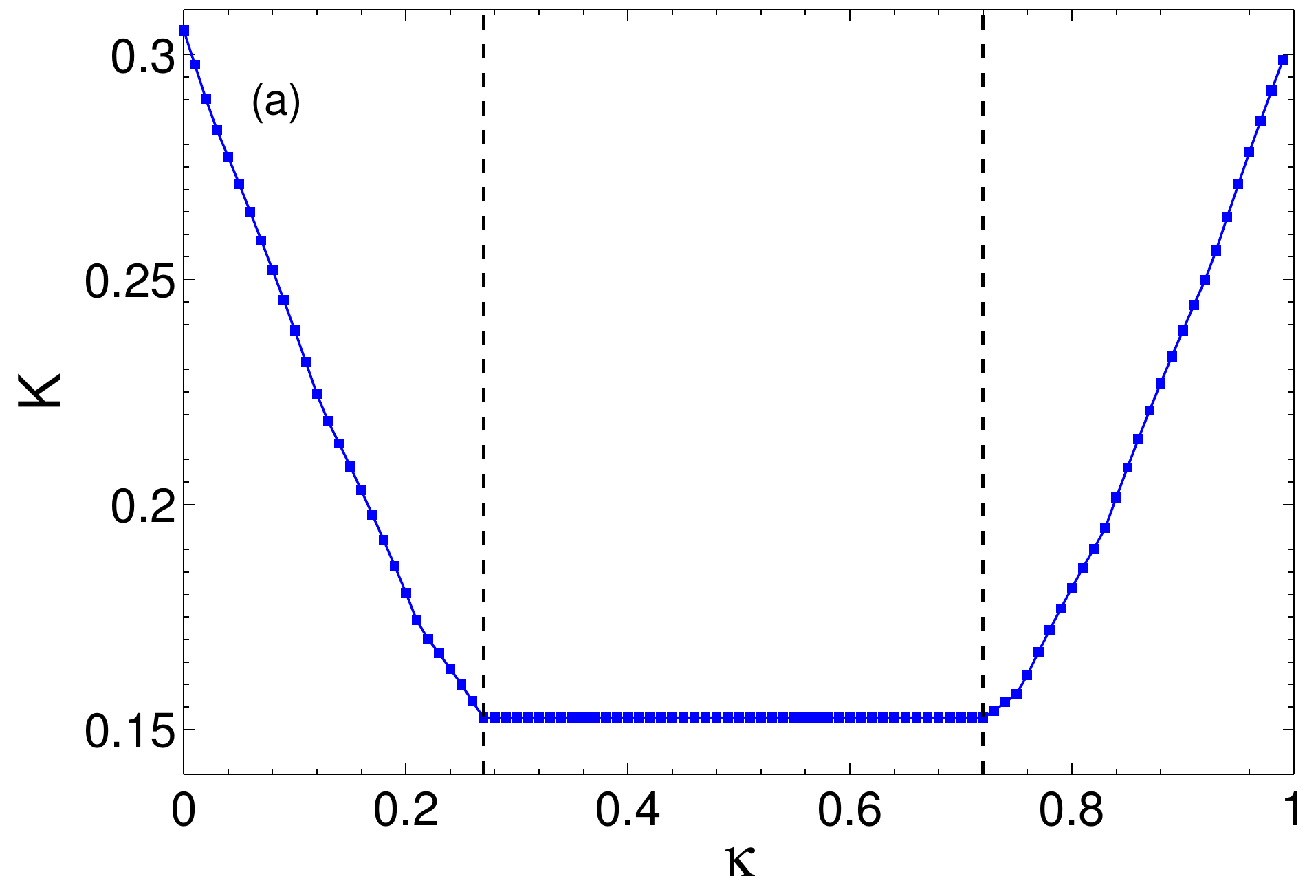}
        \includegraphics[width=6.5cm]{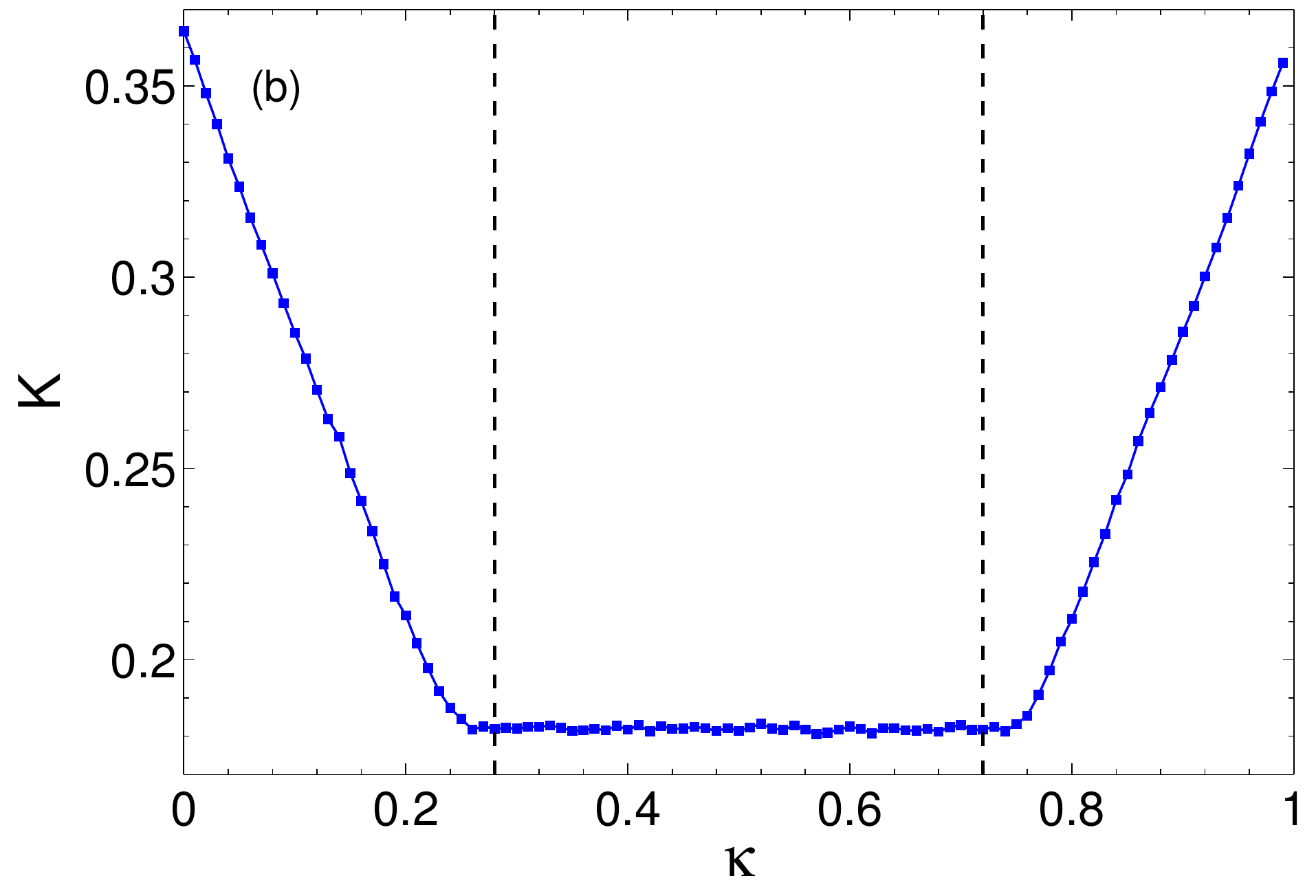}
	\caption{Kolmogorov entropy $K$ vs.\ coupling strength $\kappa$ for a system of two mutually coupled (a) Bernoulli and (b) tent maps, respectively. The vertical dashed lines indicate $\kappa_c$. The parameters are $\tau=20$, $\epsilon=0.6$, $a=1.5$ for the Bernoulli map and $a=0.4$ for the tent map, respectively.}
	\label{Fig2UnitsKolmogorovKap}
\end{figure}

The Kolmogorov entropy, computed from eq.~\eqref{EqnKolmogorov}, is shown for a pair of Bernoulli and tent maps, respectively, in Fig.~\ref{Fig2UnitsKolmogorovKap}. In the synchronized region only the $\gamma_1$ band has positive \acp{LE} which contribute to the Kolmogorov entropy. At the synchronization transition a kink in the entropy as a function of feedback strength can be seen when suddenly \acp{LE} from the other band contribute.

\begin{figure}
	\centering
        \includegraphics[width=6cm]{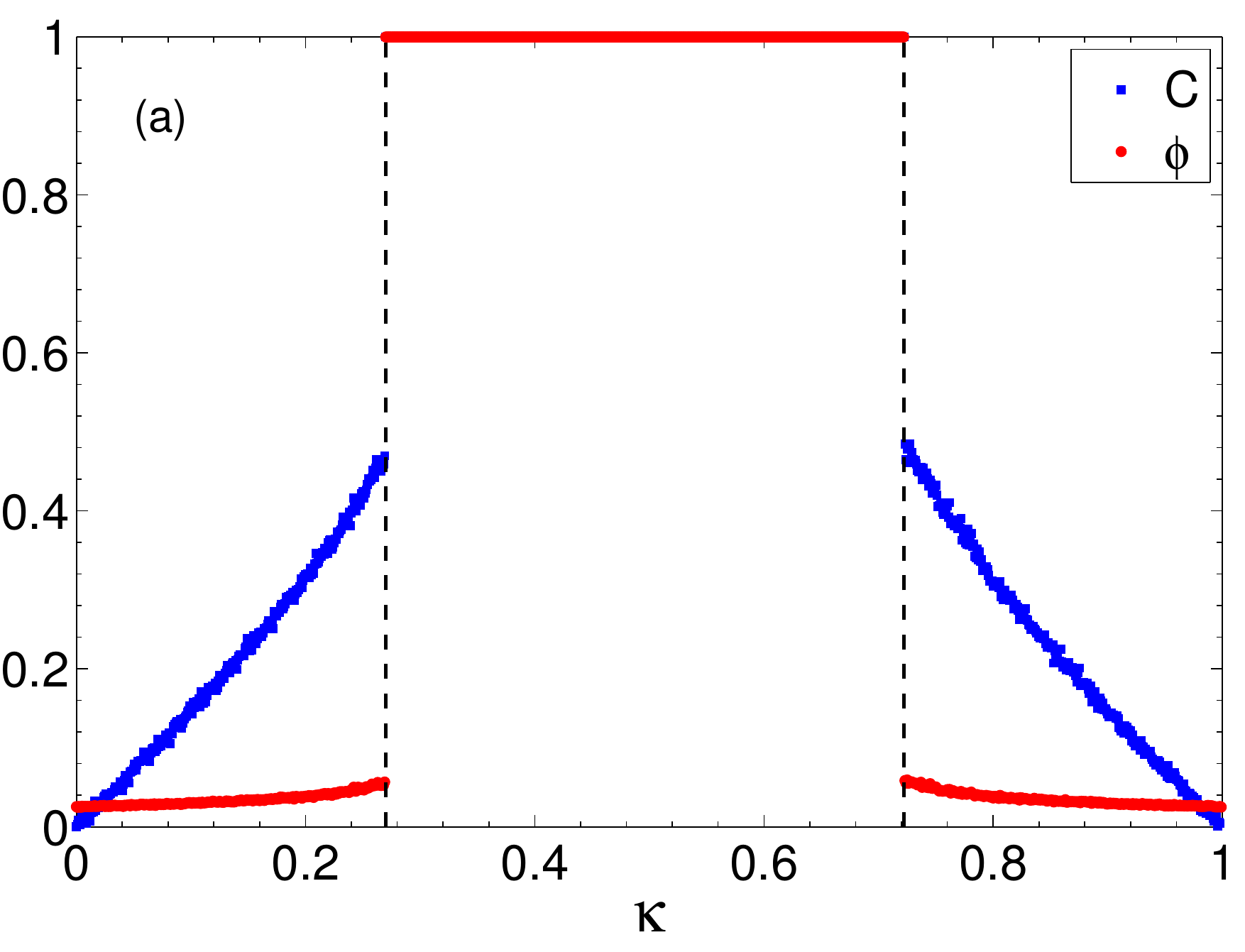}    
        \includegraphics[width=6cm]{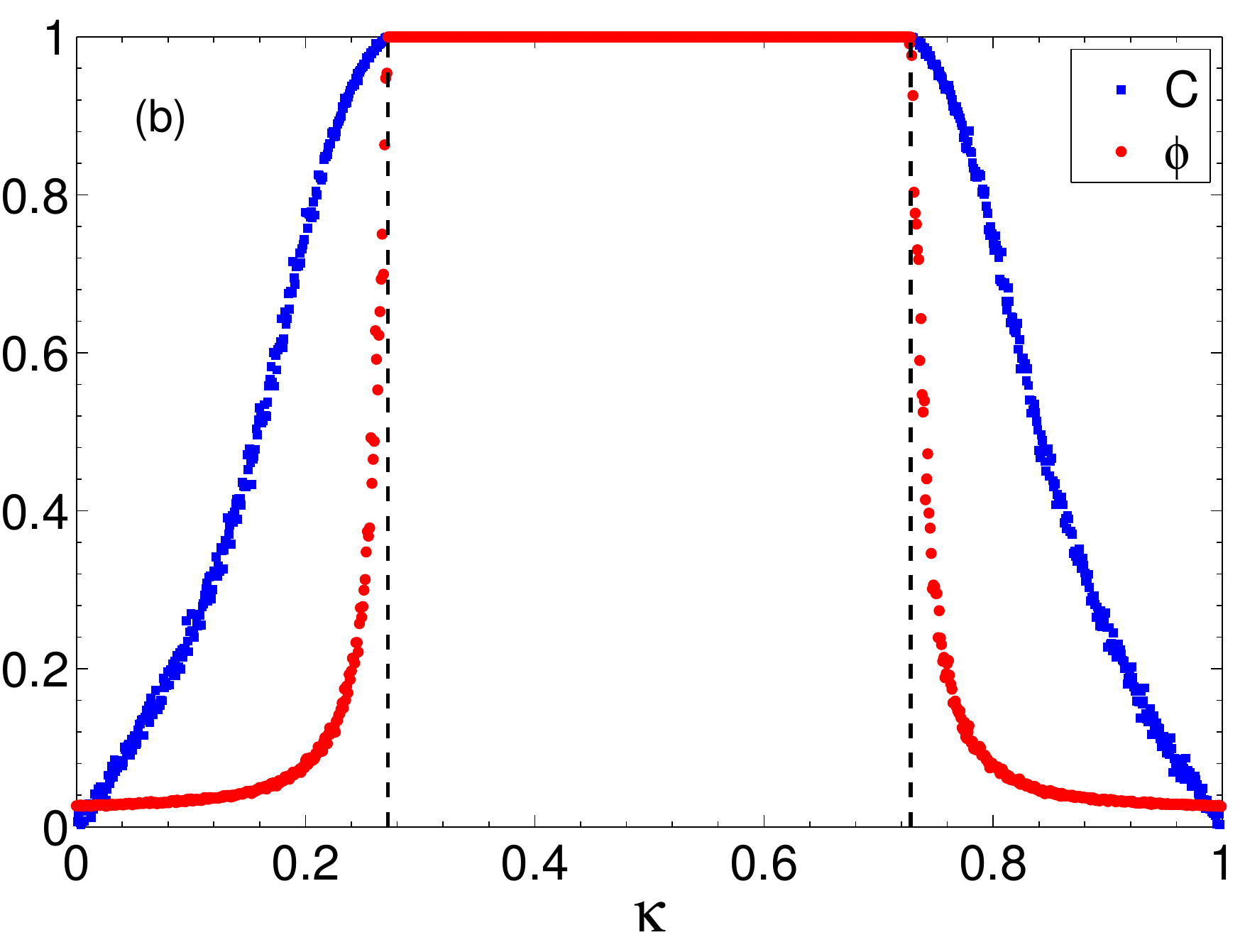}
	\caption{(Color online) Cross-correlation $C$ (blue squares) and synchronization probability $\phi$ (red dots) vs.\ coupling strength $\kappa$ for a system of two mutually coupled (a) Bernoulli and (b) tent maps, respectively. The step size is $\Delta \kappa =10^{-3}$ and the other parameters are $\tau=20$, $\epsilon=0.6$, $a=1.5$ for the Bernoulli map and $a=0.4$ for the tent map, respectively. The threshold for $\phi$ was set to $\Theta=0.01$, i.e., the trajectories were assumed to be synchronized when they were closer together than 1\% of their maximum distance.}
	\label{Fig2UnitsCorrSyncProb}
\end{figure}

The result for the attractor dimensions and the prediction entropy are very similar for Bernoulli and tent maps. But we found a qualitative difference between the two models for the cross-correlations, $C$, and the synchronization probability, $\phi$, of a system of two mutually coupled units. The synchronization probability measures the fraction of time where the two trajectories are closer than some threshold $\Theta$ \cite{Book-Llakshmanan2011}. Fig. \ref{Fig2UnitsCorrSyncProb} shows $C$ and $\phi$ with respect to the coupling strength $\kappa$ for fixed $\epsilon$. At the critical coupling $\kappa_c$, i.e., at the synchronization transition both quantities, $C$ and $\phi$, jump from a very low level to complete synchronization, $C=\phi=1$, for a system of Bernoulli maps, whereas for the tent map $C$ and $\phi$ increase continuously to $C=\phi=1$.

The numerical results indicate that the synchronization transition for the tent map is of a supercritical type.
That is, close to the transition to synchronization there is a stable trajectory close to the \ac{SM} and the dynamics is nearly synchronized. We thus observe a smooth transition to synchronization. In the Bernoulli map, on the other hand, the transition is of a subcritical type. Note that in both cases, a jump of \ac{KY} dimension is predicted due to the bands of \acp{LE}.

%

\section{Coupled lasers}

An important application of eq.~\eqref{net} is the modeling of
semiconductor lasers which are coupled by their mutual laser beams. To
a good approximation, the dynamics of the laser intensity can be
described by the Lang-Kobayashi rate equations
\cite{LangKobayashi1980}. The Lang-Kobayashi equations describe the
dynamics of a laser with delayed feedback (or delayed coupling) in
terms of a slowly varying complex electric field $E(t)$ and a
population inversion $n(t)$. For our network of coupled lasers, the
corresponding equations of Lang-Kobayashi type in dimensionless form
are given by
\begin{equation}  
  \begin{split} \dot{E}^{i}(t) & =\frac{1}{2} (1+\mathrm{i}\alpha)
    n^{i}(t)E^{i}(t)
    +\sigma\sum_{j}G_{ij} E^{j}(t-\tau)\\
    T \dot{n}^{i}(t) & =p - n^{i}(t)
    -\left[1+n^{i}(t)\right]\left|E^{i}(t)\right|^{2},
  \end{split}
  \label{eq:LK}
\end{equation}
where $E^{i}(t)$ is the envelope of the complex electric field and
$n^{i}(t)$ is the renormalized population inversion of the charge
carriers of laser $i$. The model parameters are summarized in
Table~\ref{T1}. The dimensionless delay time of $\tau=100$ translates
to a delay time of the order of magnitude $1\,$ns.
\begin{table}
  \begin{tabular}{lcl}
    \hline
    \hline
    Parameter & Symbol & Value\\
    \hline
    Linewidth enhancement factor & $\alpha$ & 4\\
    Time-scale separation of &&\\
    carrier and photon lifetimes & $T$ &
    200\\
    Injection current & $p$ & 0.1\\
    Coupling strength & $\sigma$ & 0.12\\
    Coupling delay time & $\tau$ & 100 \\
    \hline
    \hline
  \end{tabular} 
  \caption{Parameters for the simulation of the Lang-Kobayashi 
    equations. \label{T1}}
\end{table}

A network of coupled lasers modeled by the Lang-Kobayashi equations can be written in the form of eq.~\eqref{net}, where $x^{i}(t)=(n^i,\nRe{E^i},\nIm{E^i})$ is now three-dimensional and contains the real and imaginary part of the electric field $E^i$ and the charge carrier inversion $n^i$ of the $i$-th laser. A single laser is not chaotic, but the delayed feedback and/or coupling renders the system chaotic. The linear coupling function $H$ is represented by the matrix $\mathbf{H}= \left(\begin{smallmatrix}
    0 & 0 & 0\\
    0 & 1 & 0\\
    0 & 0 & 1
  \end{smallmatrix}\right)$, corresponding to all-optical coupling as
in Eq.~\eqref{eq:LK}.

We consider a pair of lasers with overall coupling strength $\sigma$
and self-feedback strength $\kappa$. That is, the coupling matrix is
given by
\begin{equation} \label{eq:1} G = \begin{bmatrix}
    \kappa & 1 -\kappa\\
    1 - \kappa & \kappa
  \end{bmatrix} \; .  
\end{equation} 
Note that in the \ac{SM}, the trajectory eq.~\eqref{sm} does not
depend on the parameter $\kappa$. Similar to the case of the maps
discussed before, the spectrum of \acp{LE} is obtained from a
Gram-Schmidt orthonormalization procedure according to Farmer
\cite{Farmer-1982}. From this spectrum we obtain the \ac{KY} dimension
using eq.~\eqref{EqnKaplanYorke}.
\begin{figure} \centering
  \includegraphics[width=\linewidth]{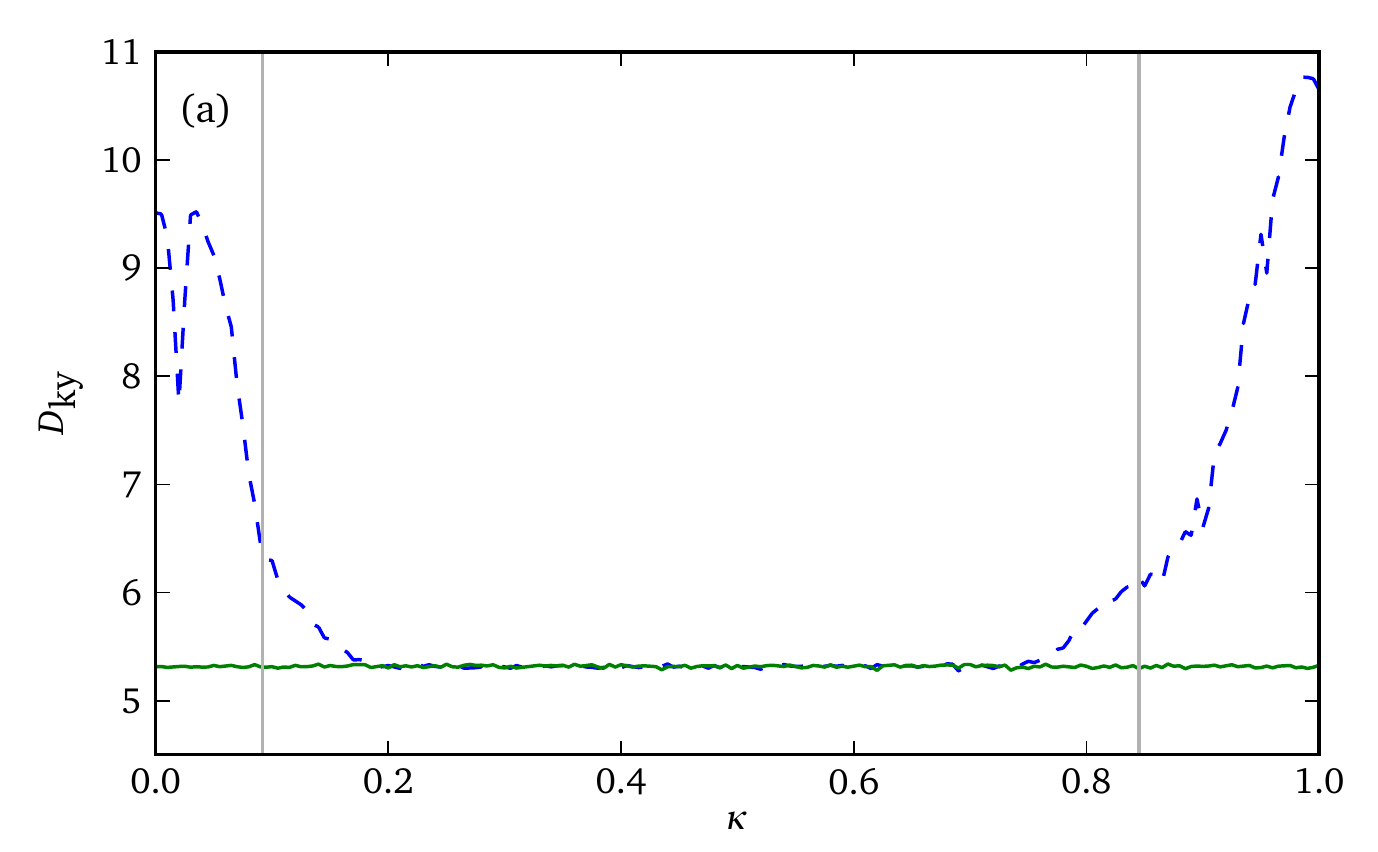}
  \includegraphics[width=\linewidth]{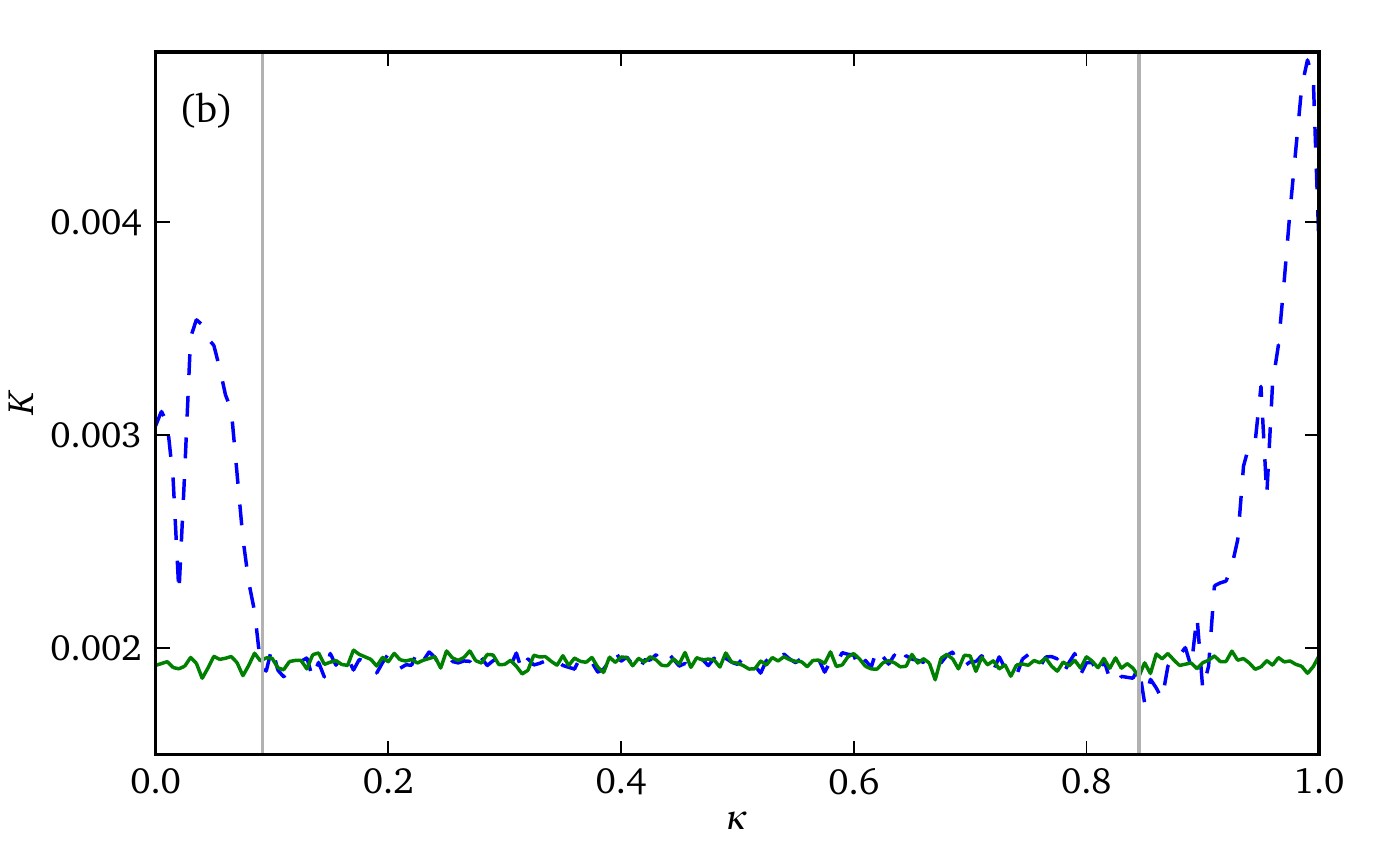}
  \caption{(Color online) (a) Kaplan-Yorke dimension $D_{KY}$ and (b)
    Kolmogorov-Sinai entropy $K$ of two coupled semiconductor lasers
    in dependence on the relative self-feedback strength $\kappa$, cf.
    Eq.~\eqref{eq:1}. The dashed blue and solid green lines are
    obtained using the complete spectrum and the using the spectrum
    inside the SM only, respectively.}
  \label{fig:ky_laser}
\end{figure}
Fig.~\ref{fig:ky_laser}(a) shows $D_{KY}$ as a function of $\kappa$.
The dashed blue curve was obtained using the complete spectrum, while
the solid green curve uses only the spectrum inside the SM. The
vertical gray lines denote the boundary of stable synchronization of
the two lasers with the given parameters; synchronization is stable
between the two lines.

Outside the synchronization region the complete spectrum has to be
used when computing the \ac{KY} dimension. However for chaos synchronization only
the spectrum in the \ac{SM}, eq.~\eqref{sm}, needs to be used. Thus,
at the transition the \ac{KY} dimension has to jump from a high
(dashed blue) to a lower value (solid green) for the \ac{SM}.

Unfortunately, we are not able to calculate the correlation dimension
of the laser rate equations. Due to the delay term which makes the
system high-dimensional, the available algorithms do not produce
reliable results, to our knowledge.

We have also calculated the Kolmogorov-Sinai entropy $K$. It is
defined as the sum of all positive \acp{LE}.
Fig.~\ref{fig:ky_laser}(b) shows the result $K(\kappa)$ for the pair
of lasers, as before. Again, the dashed blue curve was obtained using
the complete spectrum, while the solid green curve uses only the
spectrum inside the SM. At the transition, the derivative of
$K(\kappa)$ is discontinuous. In the desynchronized region, the second
band of \acp{LE} crosses zero, as shown in Fig.~\ref{FigLyapSpecEps}
and \ref{FigLyapSpecEpsTent} for the Bernoulli and the tent map
triangle, respectively, and therefore it contributes to the
Kolmogorov-Sinai entropy.

\section{Networks}

In this section, we investigate the transition to chaos synchronization in large networks. For simplicity, we consider Bernoulli networks of $N$ units with all-to-all couplings without self-feedback. The coupling matrix $G$ of eq.~\eqref{d-net} has the eigenvalues $\gamma_1=1$ and $\gamma_j=-1/(N-1), 1< j \le N$.

\begin{figure}
	\centering
	\includegraphics[width=\linewidth]{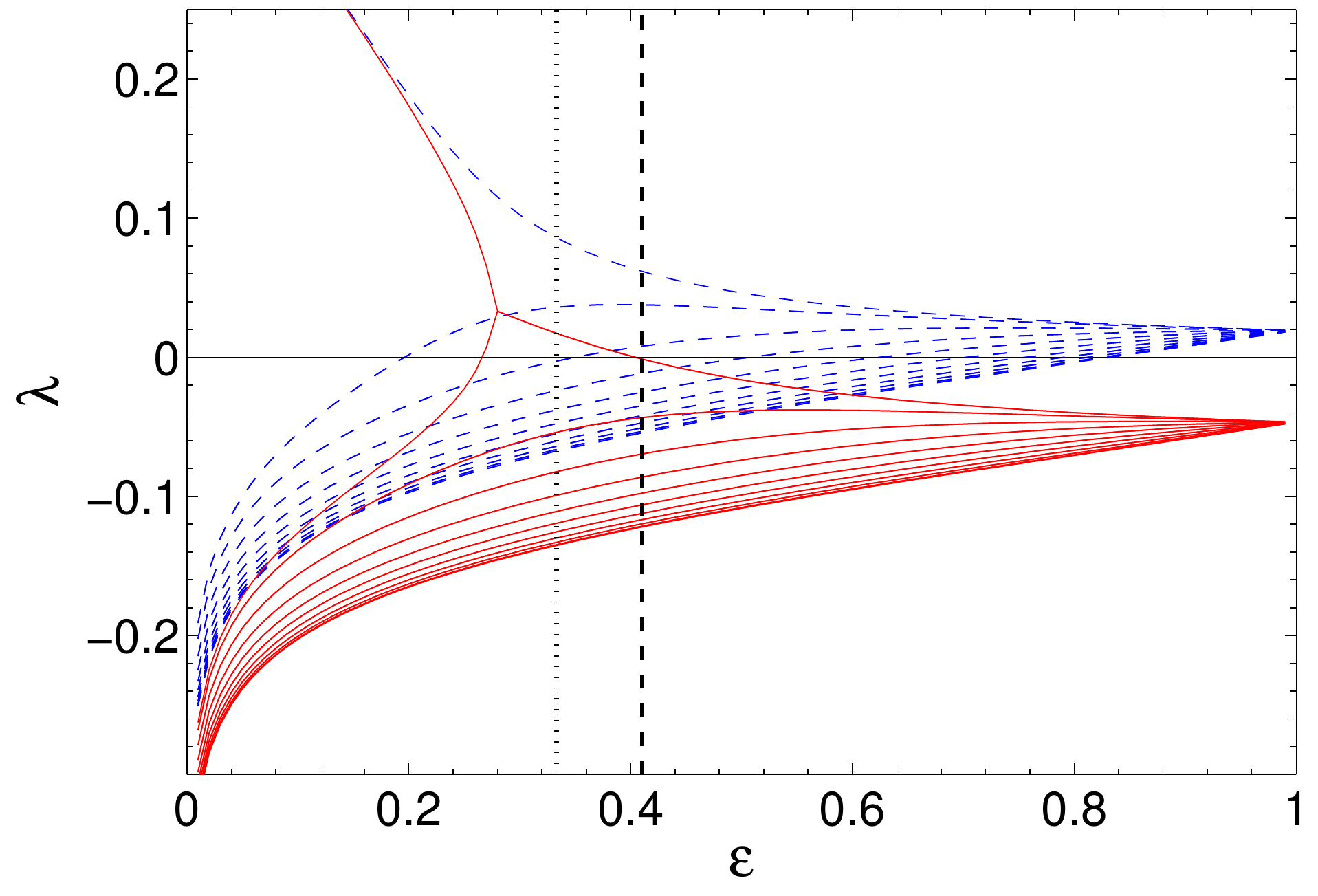}
    \caption{(Color online) Lyapunov spectra vs.\ coupling strength $\epsilon$ for a system of five all-to-all coupled Bernoulli maps with the parameters $a=1.5$ and $\tau=20$. The blue dashed lines show the longitudinal spectrum associated with $\gamma_1=1$ and the red solid lines show the transversal spectrum associated with the four-fold degenerated eigenvalue $\gamma_2=-1/4$. The vertical dotted line indicates the transition between strong and weak chaos and the vertical dashed line indicates $\epsilon_c$ where to its right the system is synchronized.}
	\label{FigLyapEpsScan}
\end{figure}

Fig.~\ref{FigLyapEpsScan} shows the spectrum of \acp{LE} as a function of $\epsilon$ for a system of five all-to-all coupled Bernoulli units. Since the eigenvalue corresponding to the transversal spectrum has the multiplicity four this spectrum is four-fold degenerated. For $\epsilon=0$, the uncoupled Bernoulli units have the LE $\lambda=\log a$. Hence the \ac{KY} dimension is $D_{KY}=N$, i.e., the full phase space. As discussed before, at the transition to synchronization only the $\gamma_1$ band contributes to $D_{KY}$, hence the \ac{KY} dimension jumps at the transition. The entropy has a kink, since at the transition the $N-1$ many $\gamma_j$ bands do not contribute any more to the entropy.

We consider the limit of large delay times, $\tau \to \infty$. In this
limit, using eq. \eqref{PolynomialEqn}, we find the critical value at
which the transition to synchronization occurs to be $\epsilon_c =
(a-1)/(a(1- |\gamma_j|))$. For a coupling strength $\epsilon < 1 -1/a
$ the system is in the regime of strong chaos otherwise in the regime
of weak chaos \cite{PhysRevLett.107.234102}. In general the transition
from strong to weak chaos does not coincide with the synchronization
transition which usually occurs at larger values $\epsilon_c \geq 1
-1/a$, see eq.~\eqref{gap}. Only in the limit of $N \to \infty$ both
transitions fall together. The transition from strong to weak chaos is
defined by the change in sign of the so-called local Lyapunov exponent
\cite{PhysRevLett.107.234102}. For strong chaos it is positive and for
weak chaos negative.

For strong chaos the maximum \ac{LE} is of order one whereas for weak chaos it scales as $1/\tau$. Since each mode of the network has $\tau$ many \acp{LE} of the order of $1/\tau$, the \ac{KY} dimension increases linearly with the delay time $\tau$ whereas the Kolmogorov entropy $K$ is independent of $\tau$.
For a large enough coupling strength, $\epsilon$, such that the system is synchronized, the dimension is determined solely by the \ac{SM} and therefore the dimension is independent of the number of units, $N$, and cannot exceed a value larger than the delay time of a single unit.

\begin{figure}
	\centering
        \includegraphics[width=7cm]{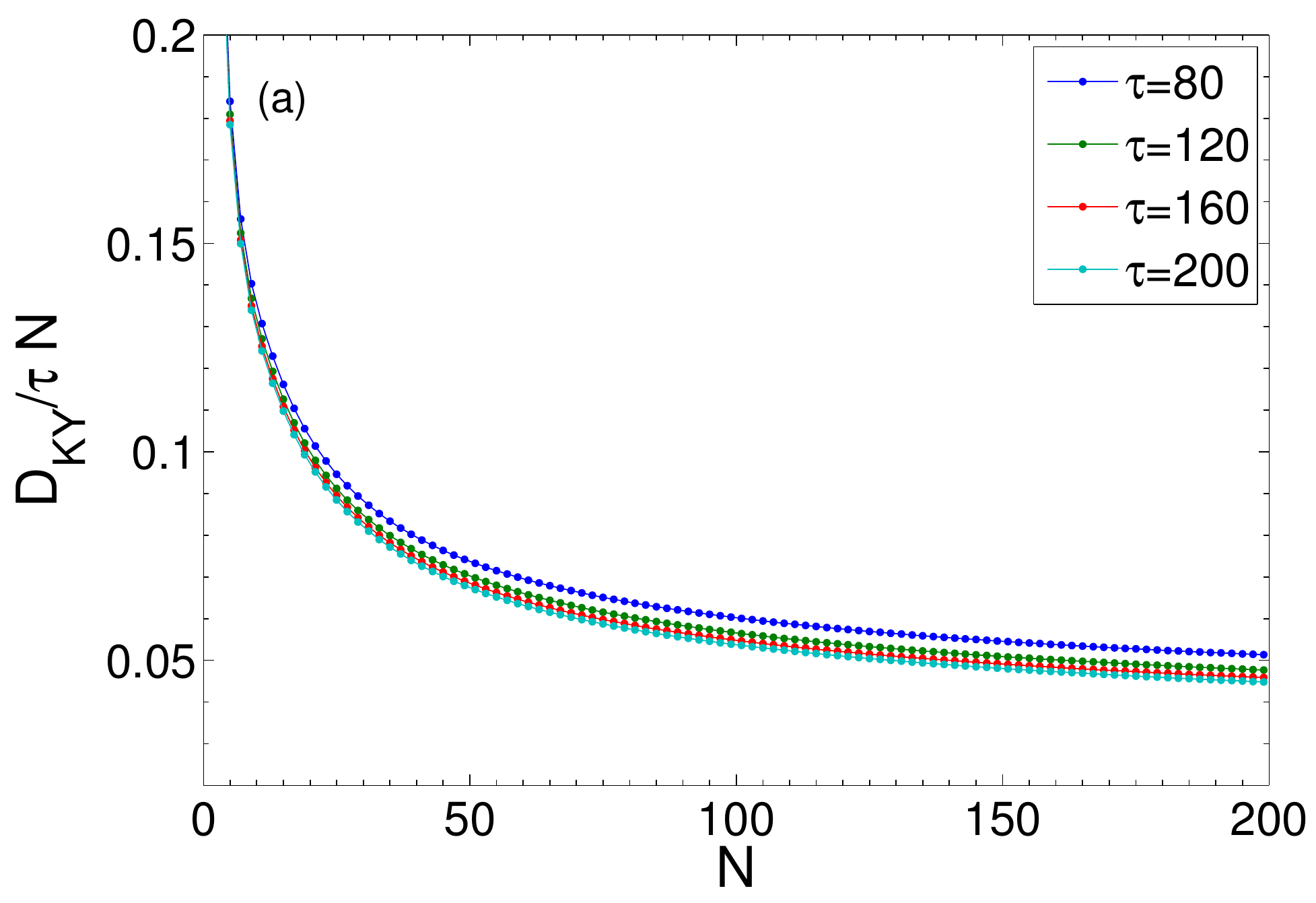}
        \includegraphics[width=7cm]{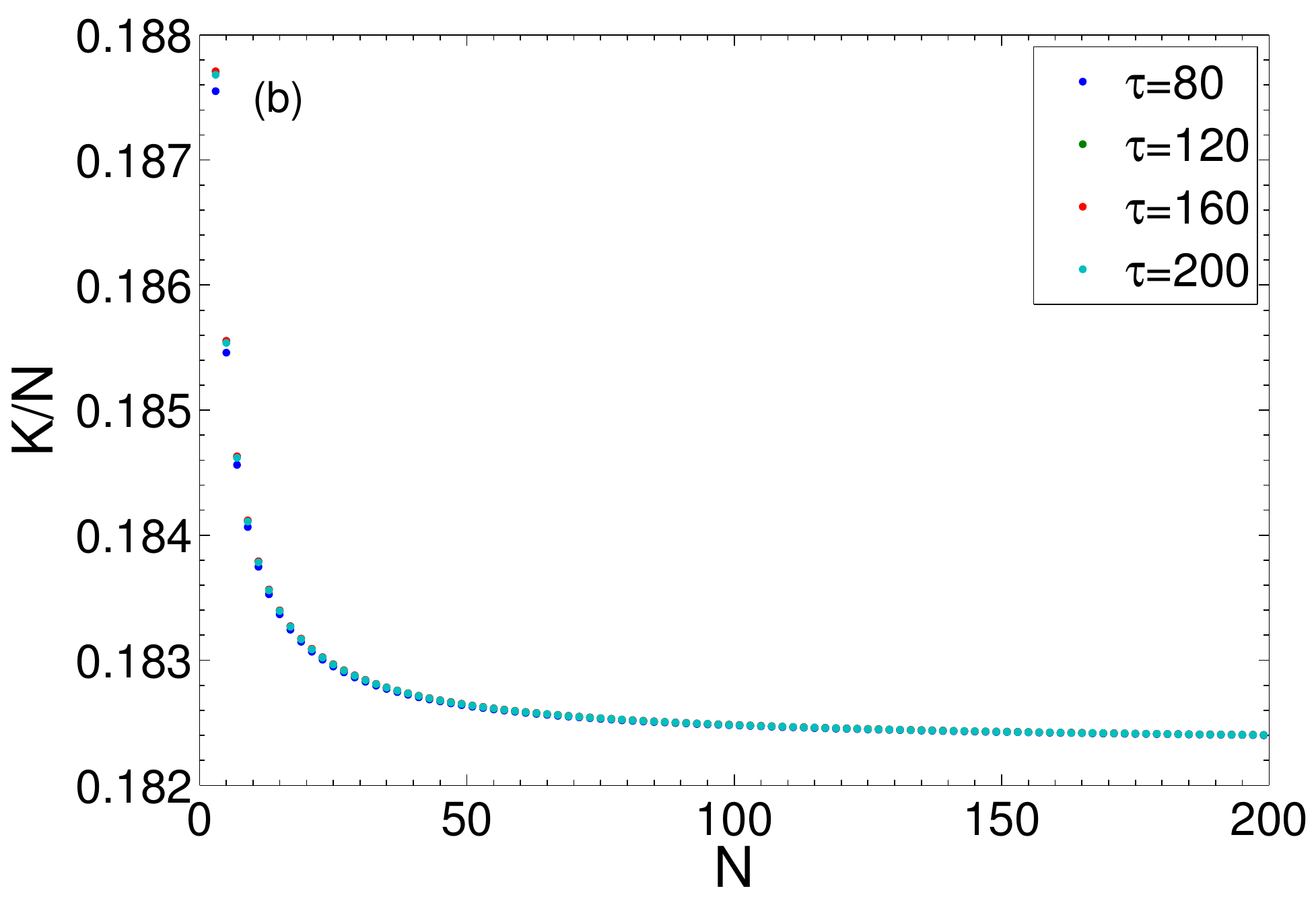}
    \caption{(Color online) (a) KY dimension $D_{KY}$ and (b) Kolmogorov entropy $K$ with respect to the number of all-to-all coupled Bernoulli maps $N$ for different delay times $\tau$. It starts from $\tau=80$ (uppermost curve) and increases in steps of $\Delta \tau=40$ up to $\tau=200$ (lowermost curve). The other parameters are $a=1.5$ and $\epsilon=0.2$.}
	\label{FigNScanTauScanStrongChaos}
\end{figure}

Fig.~\ref{FigNScanTauScanStrongChaos} shows the \ac{KY} dimension and the Kolmogorov entropy as a function of the system size, $N$, in the regime of strong chaos for different delay times $\tau$. It is clearly visible that for large delay times the different plots of $D_{KY}/\tau N$ corresponding to different delays nearly coincide. Thus the \ac{KY} dimension scales linearly with $\tau$ whereas the Kolmogorov entropy is independent of $\tau$. Both, entropy as well as dimension increase with system size $N$ and it seems that for large $N$ both quantities scale linearly with $N$.

As discussed before, the transition to synchronization occurs at $\epsilon_c$ which depends on the number of units in the network. In the limit of large delay times $\tau \to \infty$ we find $\epsilon_c = (a-1)/(a(1-|\gamma_j|))$. Thus $\epsilon_c$ decreases monotonically towards $\epsilon = 1 -1/a$ for $N \to \infty$.
At the transition only the $\gamma_1$ band contributes to $D_{KY}$ and hence the \ac{KY} dimension jumps.

\begin{figure}
	\centering
        \includegraphics[width=7cm]{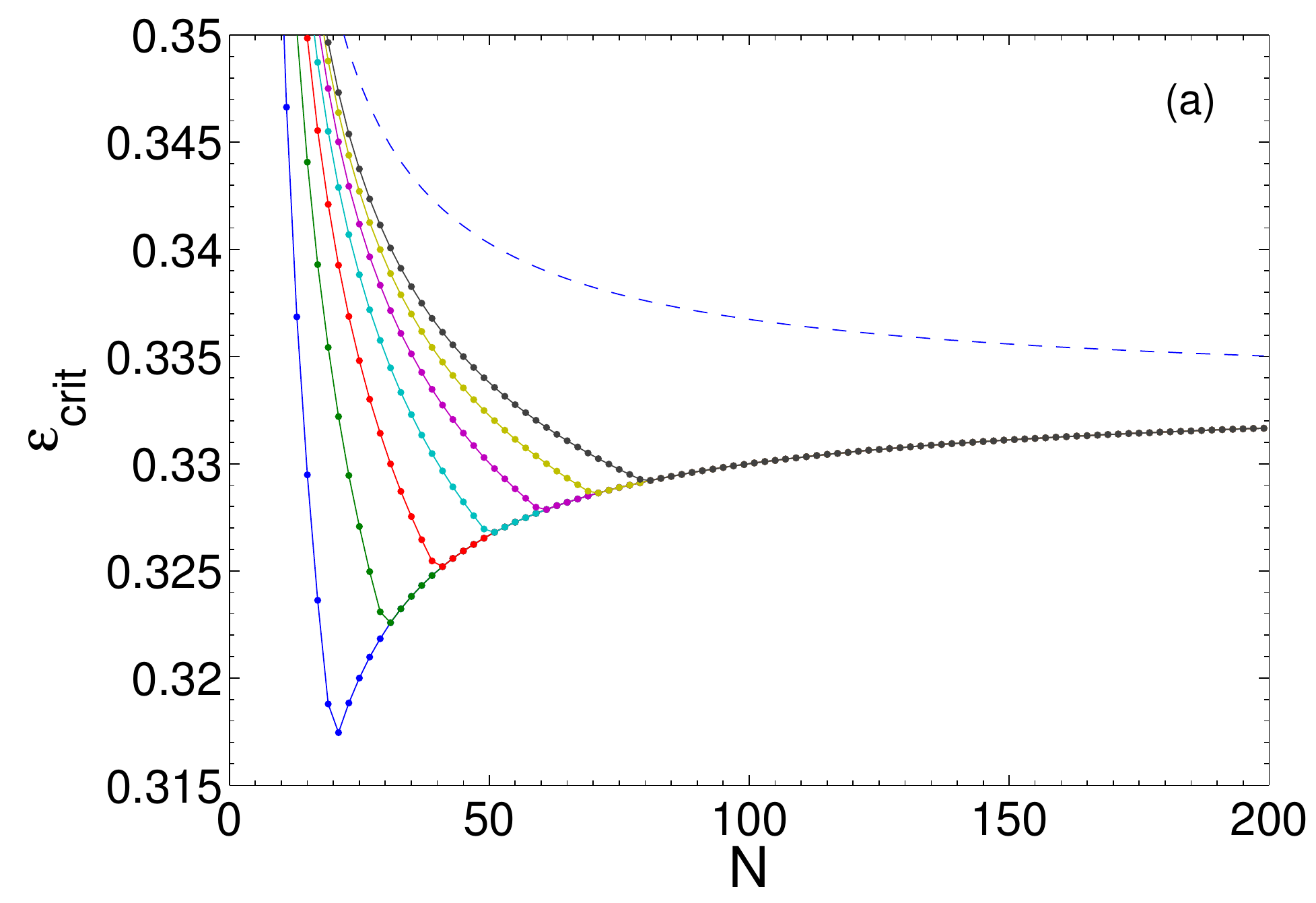}
        \includegraphics[width=7cm]{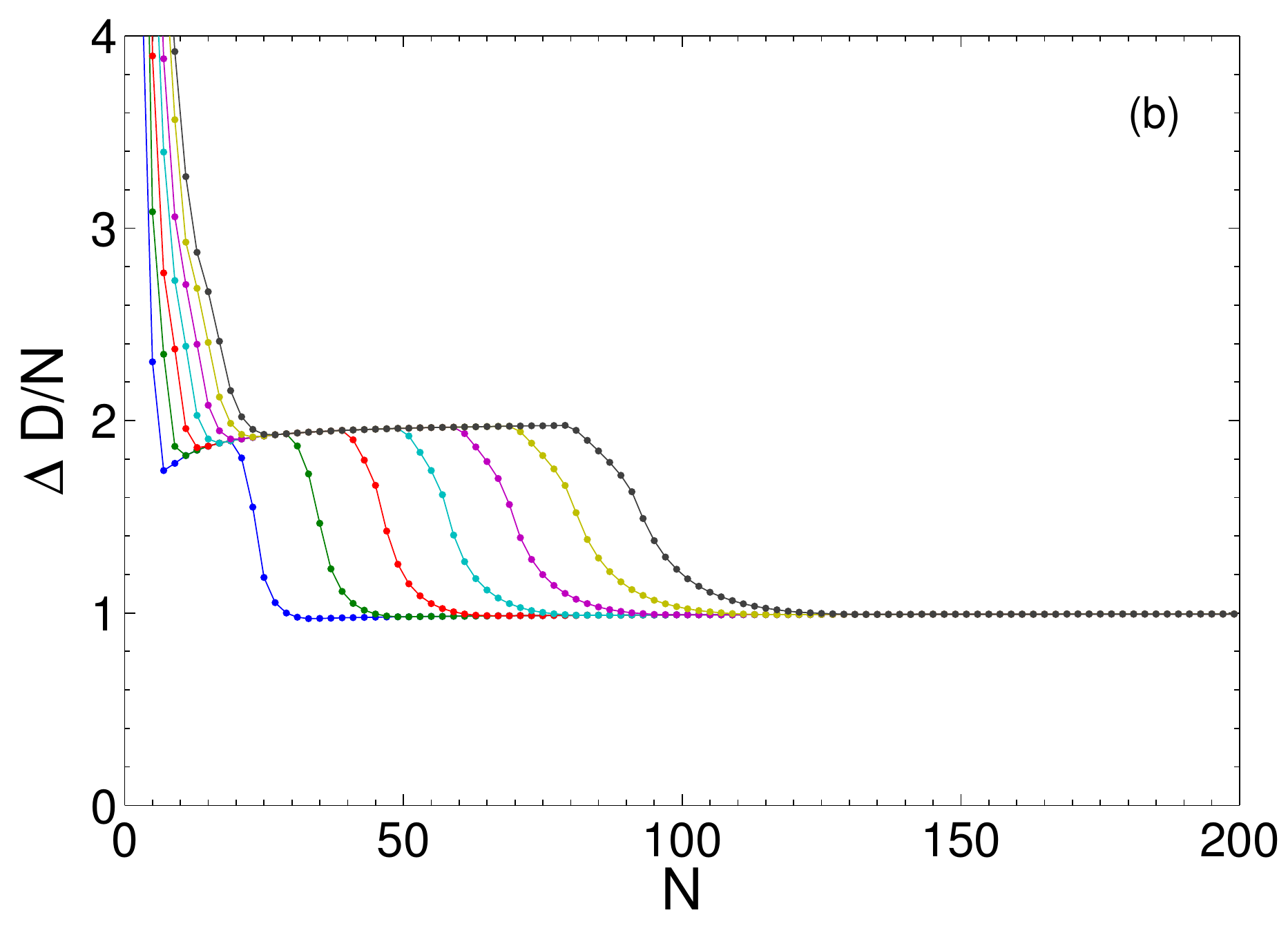}
        \includegraphics[width=7cm]{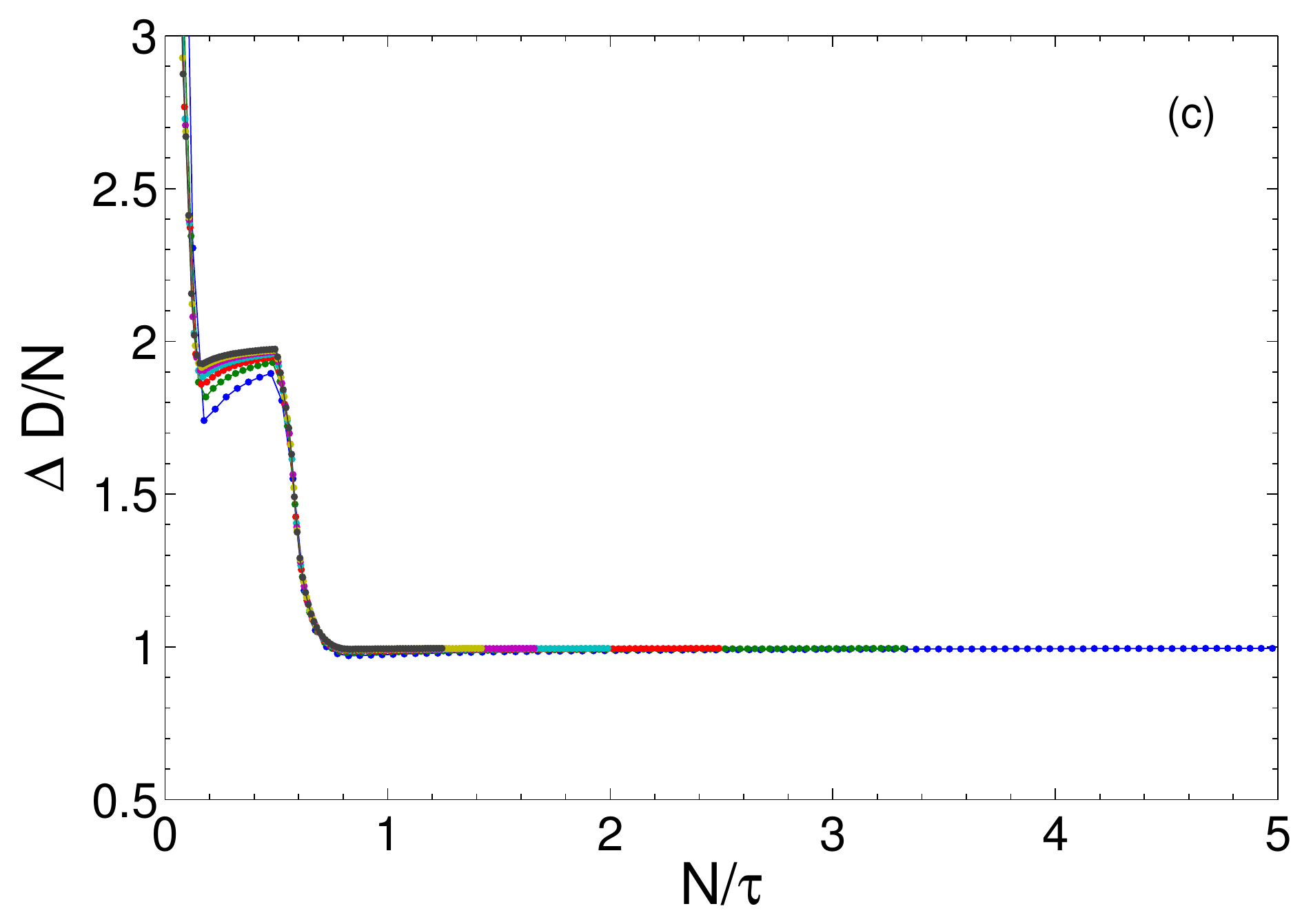}
        \caption{(Color online) (a) Critical coupling strength $\epsilon_c$ and (b,c) jump of the KY dimension $\Delta D_{KY}$ with
          respect to the system size $N$ of all-to-all coupled
          Bernoulli maps. The different curves correspond to different
          delay times, $\tau$. Lowest curve is for $\tau=40$ and
          increases by steps of $\Delta \tau = 20$ up to $\tau=160$
          (upper curve). For $\epsilon_c$ the theoretical value in the
          limit of $\tau \to \infty $ is plotted as well (upper dashed
          line). The parameter of the Bernoulli map is $a=1.5$.
 }
	\label{FigDimJump}
\end{figure}

Fig.~\ref{FigDimJump} shows the jump of the \ac{KY} dimension at the
transition as well as the critical coupling strength $\epsilon_c$ at
which the jump occurs as a function of network size $N$ for different
delay times. Since $\epsilon_c$ depends on $N$, the jump has a
nonmonotonic behavior. For large enough network sizes the jump $\Delta
D$ scales linearly with the number of units. 
The slope of this linear relation is approximately two for small $N$, $\Delta D/N \approx 2$, and approximately one for large $N$, $\Delta D/N \approx 1$.
The transition between the two different slopes is related to the dip in the
critical coupling strength $\epsilon_c(N)$ which is due to the
fact that the system is not yet in the limit $\tau \to \infty$.

The slope  $\Delta D/N$ depends on the order of limits. If we take the limit $\tau
\to \infty$ first the jump $\Delta D$ scales linearly with the number of units $N$
with a slope of two. If we however take the limit $N \to \infty$ first
it scales with a slope of one for any value of $\tau$.

In Fig.~\ref{FigDimJump} (c) the same data as in panel (b) is plotted
as $\Delta D / N$ versus $N/\tau$ leading to data-collapse. This shows
that the jump in the Kaplan-Yorke dimension is in the limit of large
$\tau$ determined by the scaling law
\begin{equation*}
   \Delta D \approx N \psi(N/\tau) \:,
\end{equation*}
where $\psi$ is the scaling function depicted in panel (c).

\section{Summary and discussion}

Networks of identical nonlinear units can synchronize to a common chaotic trajectory. Although the time-delay of the coupling can be very large, the units can synchronize without any time shift. We investigated the transition to chaos synchronization in the limit of very large delay times. General arguments about the dynamic of such a network predict a jump of the \ac{KY} dimension of the chaotic attractor when the network synchronizes. In addition, the Kolmogorov prediction entropy should show a discontinuous slope.

We tested these general predictions for networks of iterated maps.
For Bernoulli maps, our numerical results show a clear discontinuous behavior of the attractor dimension. The \ac{KY} as well as the correlation dimension jump to a low value when the network synchronizes. For tent maps, the numerical results of the correlation dimension are not so clear due to large statistical fluctuations caused by limited computational power. Nevertheless, our results indicate a jump in the attractor dimension, too. In both cases the prediction entropy shows a discontinuous slope.

The \ac{KY} dimension was also calculated for the rate equation of semiconductor lasers. Again, our general arguments give a discontinuous \ac{KY} dimension at the transition. Unfortunately, we were not able to calculate the correlation dimension for this case.

For Bernoulli networks we numerically calculated the \ac{KY} dimension as a function of system size $N$ and delay time $\tau$ in the region of strong chaos. The dimension scales with $N \tau$. The jump of the dimension at the transition scales with $N$, as well.

Our results show that one has to use the standard definition of the \ac{KY} dimension, eq.~\eqref{EqnKaplanYorke}, with care. If we use eq.~\eqref{EqnKaplanYorke} with all possible \acp{LE} we obtain a wrong result. Here we argued that not all \acp{LE} contribute to eq.~\eqref{EqnKaplanYorke} in case of synchronization. This argument predicts a jump in the attractor dimension, in agreement with our numerical results of the correlation dimension. However, in general it may not be obvious which \acp{LE} contribute to the \ac{KY} dimension. For example, if we distort the synchronization manifold by a nonlinear transformation of one unit the dimension does not change but we do not know which \ac{LE} have to be omitted in this case. Also, in the case of generalized synchronization the dynamics is restricted to a low dimensional manifold which rules out the majority of \acp{LE}. But again, we do not know in advance which \acp{LE} have to be omitted from eq.~\eqref{EqnKaplanYorke}.

Our argument of omitting bands of negative \acp{LE} relies on the fact that the dynamics is restricted to the \ac{SM}. However, a tiny detuning of the nonlinear units leads to imperfect synchronization and this argument is no longer valid. Thus, the attractor dimension should immediately jump to a high value for any amount of detuning. We have tried to calculate the dimension in the limit of zero detuning but our numerical results did not allow a conclusive statement.

\bibliographystyle{apsrev4-1}


%

\begin{acronym}[MSF]
  \acro{SM}{synchronization manifold} \acro{LE}{Lyapunov exponent}
  \acro{SM}{synchronization manifold} \acro{KY}{Kaplan-Yorke}
\end{acronym}

\end{document}